# Engineering Three-Dimensional (3D) Out-of-Plane Graphene Edge Sites for Highly Selective Two-Electron Oxygen Reduction Electrocatalysis

*Daniel San Roman*[1^], *Dilip Krishnamurthy*[2^], *Raghav Garg*[1], *Hasnain Hafiz*[2], *Michael Lamparski*,[4] *Noel T. Nuhfer*[1], *Vincent Meunier*,[4] *Venkatasubramanian Viswanathan*[2*] and *Tzahi Cohen-Karni*[1,3*]

[1] *Department of Material Science and Engineering, Carnegie Mellon University, Pittsburgh, Pennsylvania, 15213, USA.*

[2] *Department of Mechanical Engineering, Carnegie Mellon University, Pittsburgh, Pennsylvania, 15213, USA.*

[3] *Department of Biomedical Engineering, Carnegie Mellon University, Pittsburgh, Pennsylvania, 15213, USA.*

[4] *Department of Physics, Applied Physics, and Astronomy, Rensselaer Polytechnic Institute, Troy, New York 12180, USA*

^ *Equal Contribution*

* *Corresponding author*

## Abstract

Selective two-electron oxygen reduction reaction (ORR) offers a promising route for hydrogen peroxide synthesis, and defective $sp^2$ carbon-based materials are attractive, low-cost electrocatalysts for this process. However, due to a wide range of possible defect structures formed during material synthesis, identification and fabrication of precise active sites remain a challenge. Here, we report a graphene edge-based electrocatalyst for two-electron ORR – nanowire-templated three-dimensional fuzzy graphene (NT-3DFG). NT-3DFG exhibits excellent efficiency (onset potential of 0.79 ± 0.01 V versus RHE), selectivity (93 ± 3 % $H_2O_2$), and tunable ORR activity as a function of graphene edge site density. Using spectroscopic surface characterization and density functional theory calculations, we find that NT-3DFG edge sites are readily functionalized by carbonyl (C=O) and hydroxyl (C–OH) groups under alkaline ORR conditions. Our calculations indicate that multiple site configurations at both armchair and zigzag edges may achieve a local coordination environment that allows selective, two-electron ORR. We derive a general geometric descriptor based on the local coordination environment that provides activity predictions of graphene surface sites within ~0.1 V of computed values. Herein we combine synthesis, spectroscopy, and simulations to improve active site characterization and accelerate carbon-based electrocatalyst discovery.

## Introduction

Hydrogen peroxide ($H_2O_2$) is a versatile and environmentally compatible oxidizing agent, and the relative safety and simplicity of its use has led to a number of applications in paper and textile manufacturing and wastewater treatment systems.[1] Currently, $H_2O_2$ is produced through an energy-intensive anthraquinone oxidation process that requires large-scale infrastructure and generates considerable amounts of waste.[1] On-site electrochemical production where oxygen reduction reactions (ORR) directly yield $H_2O_2$ offers an attractive, alternative synthesis route,[2, 3] and recent interest in the sustainable electrochemical production of $H_2O_2$ has sparked a search for low-cost and selective electrocatalysts.[4-9] Noble metal alloys, such as Pd–Au and Pt–Hg, exhibit small overpotentials for ORR and high $H_2O_2$ selectivity,[4, 5] but their cost and scarcity impede widespread adoption. Metal-free carbon-based materials (e.g., oxidized carbon nanotubes[6], reduced graphene oxide[7], and edge-rich carbon nanostructures[8]) have shown promise as selective $H_2O_2$ electrocatalysts. However, the disordered nature of pyrolysis-based synthesis and defect-forming oxidative treatments lead to difficulty in describing the nature of the active sites and fabrication of model catalysts with a specified number or type of active sites. Additionally, the state of the surface under reaction conditions may differ vastly from pristine conditions,[10] and given equilibrium through water discharge reactions, numerous possible edge site terminations (-H, -OH, or -O) may be active.[11] Generating controlled catalytic sites in $sp^2$-hybridized carbon structures with known defects and morphologies through careful synthesis of graphene-based materials can systematically address this challenge.[12] This approach decreases the phase space of active sites and allows the determination of electrocatalytic activity and selectivity using density functional theory (DFT) calculations.[13, 14]

Here we report a novel graphene-based hybrid nanomaterial – nanowire-templated out-of-plane three-dimensional fuzzy graphene[15] (NT-3DFG) – and through tunable synthesis procedures, we demonstrate precise control over the size and density of out-of-plane graphene flakes and edges. By varying the density of out-of-plane graphene flakes, NT-3DFG electrodes displayed tunable ORR active site density and thus ORR activity. NT-3DFG optimized for high densities of graphene edges showed superior performance as a $H_2O_2$ electrocatalyst with excellent efficiency (onset potential of 0.79 ± 0.01 V versus RHE) and selectivity (93 ± 3 % $H_2O_2$). We constructed surface Pourbaix diagrams using data from DFT calculations to probe the termination groups for the graphene edge sites, both armchair (AC) and zig-zag (ZZ), as a function of reaction conditions (pH and electrode potential). First principle computations with parallel surface characterization based on X-ray photoemission spectroscopy (XPS) and electron energy loss spectroscopy (EELS) suggest that ZZ edge sites are saturated by high coverage carbonyl (C=O) groups while AC edges may exhibit both hydroxyl (C–OH) and carbonyl groups. We explored a range of possible active sites to account for kinetically stable sites in addition to thermodynamically stable configurations. Our calculations suggest that the oxygenation process significantly alters the surface energies of graphene edges and that zigzag edge sites with high carbonyl group coverage as well as other edge configurations with a similar local coordination environment may enable selective, two-electron ORR. A geometric descriptor for $H_2O_2$ activity was derived on the basis of the active site coordination environment, which enables activity predictions of graphene surface sites with high degree of consistency (within ≈0.1 V) with DFT calculations.

## Results

**Controlling catalyst morphology.** To better understand the contribution of graphene edges to ORR in the $sp^2$-hybridized carbon systems of interest here, the number of out-of-plane graphene edges in catalyst electrodes was controllably varied by tuning the synthesis temperature and time. This study leveraged a highly-tunable plasma enhanced chemical vapor deposition (PECVD) synthesis process to achieve nanowire-templated three-dimensional fuzzy graphene[15] (NT-3DFG) with varying graphene flake densities and size (**Figure 1.a**). Scanning electron microscope (SEM) imaging (**Figure 1.b-d**) revealed increasing NT-3DFG flake size (685 ± 60, 975 ± 115, and 1150 ± 85 nm NT-3DFG diameters) and densities (430 ± 60, 2990 ± 355, and 3300 ± 530 $\mu m^{-2}$) for samples synthesized at 700, 900, and 1100 °C for 30 min, respectively (**Figure 1.e,f**). These observations can be attributed to increasing nucleation and growth rates of graphene flakes through PECVD.[16] NT-3DFG growth follows Arrhenius kinetics with the growth activation energy ($E_A$) calculated as 0.16 ± 0.01 eV (**Figure S1**). The calculated activation energy is lower than literature reported activation energy for vertical graphene nanosheets ($E_A$ = 0.57 eV),[17] suggesting that the faster growth rate in our synthesis protocol might be responsible for increased flake size and density.

To simulate structural growth of NT-3DFG as a function of temperature, we implemented a computational algorithm based on the general method of diffusion-limited aggregation (DLA)[18] as a series of single-particle nucleation events, as illustrated in **Figure S2** and discussed in previous work.[15] Briefly, a successful nucleation occurs when an introduced carbon dimer particle (beginning at point A' and operating under Brownian motion) approaches close enough (to point C) to meet a defined criterion for nucleation based on a parameter, $n_{touch}$, that acts as a surrogate for temperature (model parameters and nucleation conditions are provided in Methods). In the proposed model, an increased $n_{touch}$ corresponds to a greater temperature, as it represents the increased kinetic energy of the particle. This method has been used in the modeling of surface roughness evolution and on-surface nanofilaments growth[19] and does not explicitly use chemical information made during bonding. DLA is typically characterized by self-similarity and properties of a fractal nature.[18] The computations performed here result primarily in neatly filled planes with few defects for all temperature conditions (**Figure 1g**). Varying synthesis temperature likely alters the time scale between nucleation events (see **Figure S1**), and therefore we do not expect simulations to represent equal intervals of time. The algorithm is found to produce significantly more densely-packed planes for greater values of $n_{touch}$, which provides qualitative agreement with experimental outcomes in terms of measured flake densities. Additionally, NT-3DFG synthesis was performed at 1100 °C for 60 and 120 min to further increase flake size and density.[15] SEM images showed similar surface topologies at extended growth times, while NT-3DFG diameter increased to 1885 ± 215 and 4230 ± 640 nm for samples synthesized at 1100 °C for 60 and 120 min, respectively (**Figure S3**).

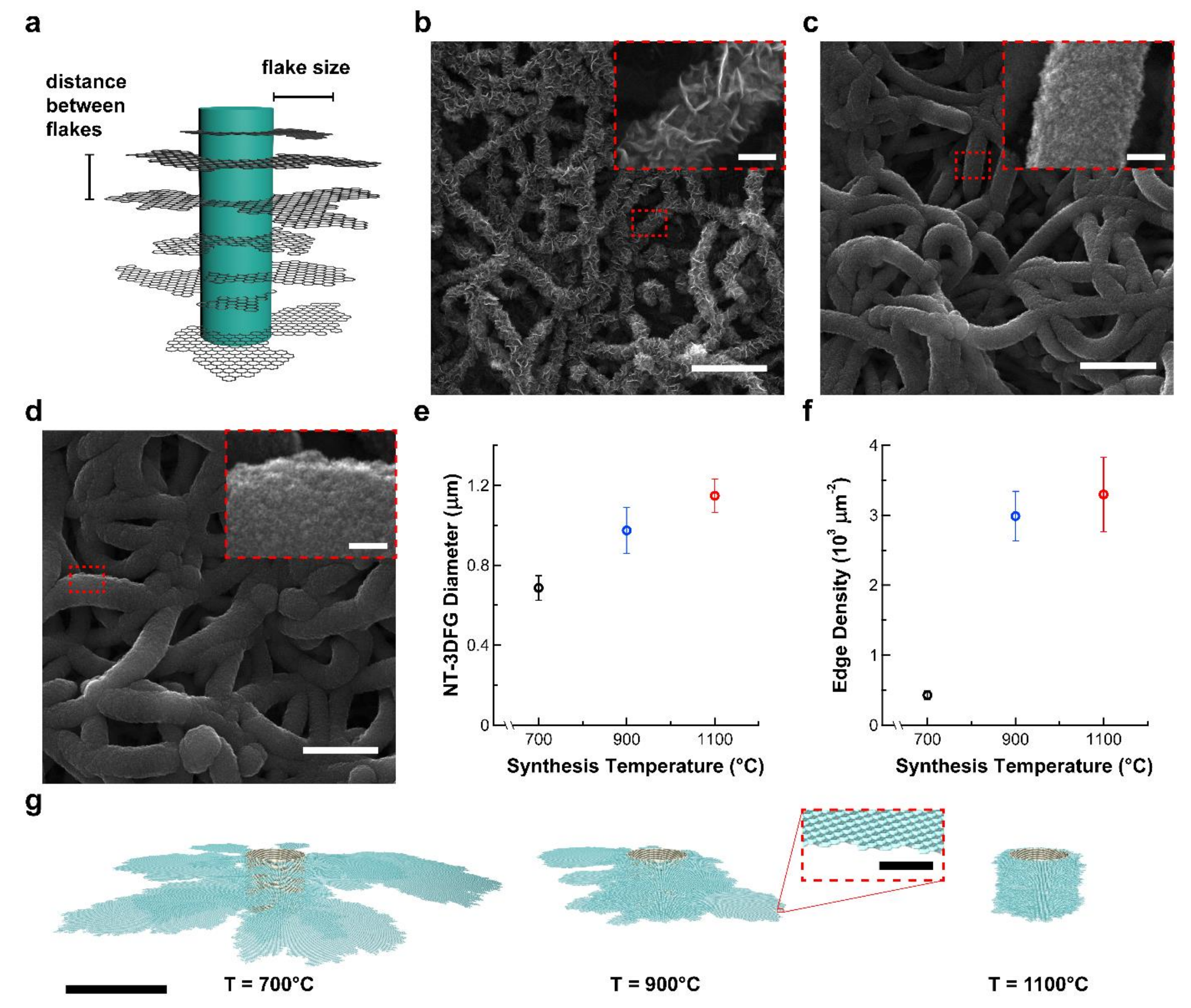

**Figure 1. NT-3DFG flake size and edge density tunability. (a)** Schematic representing NT-3DFG flakes. **(b-d)** Representative SEM images of NT-3DFG synthesized for 30 minutes at 700, 900, and 1100 °C, respectively. Scale bars: 3 $\mu$m. Insets represented by red dashed boxes show out-of-plane graphene on SiNW. Inset scale bars: 300 nm **(e)** NT-3DFG diameter and **(f)** edge density as a function of synthesis temperature. Results presented as mean ± standard deviation (SD) with sample size (*n)* of 3. **(g)** DLA-based growth model simulations of NT-3DFG for each synthesis temperature after 600,000 nucleation events. Simulations do not represent equal time intervals. Scale bar: 50 nm. Inset scale bar: 1 nm.

Across all synthesis conditions, characteristic Raman D, G, and 2D peaks for NT-3DFG were analyzed to corroborate the presence of graphene (**Figure 2.a,d** and **Table S1**).[20] The emergence of the D peak, at ~ 1335 $cm^{-1}$, and the D' peak, as a shoulder to the G peak, is due to breaks in translational symmetry due to the presence of 3DFG edges, as evident in SEM images (**Figure 1.b-d** & **Figure S3.a-c**). All samples exhibited a saturation of the G peak position near ~1580 $cm^{-1}$ and change in position of the G peak as function of excitation wavelength ($Disp_G$) < 0.10 $cm^{-1}$ $nm^{-1}$, which indicates the significant presence of aromatic $sp^2$ rings and absence of large structural defects.[20-22] The full width half maximum of the G peak ($FWHM_G$) increased (40.8 ± 2.2 $cm^{-1}$, 47.2 ± 1.6, and 49.0 ± 1.4 $cm^{-1}$) as NT-3DFG synthesis temperature (700 °C, 900 °C, and 1100 °C, respectively) increased, indicating an increase in number of edges and decrease in average flake size (**Figure 2.b.I**).[23] The rise (2.34 ± 0.52 to 3.59 ± 0.15) in the intensity ratio of the D and G peak ($I_D$ $I_G^{-1}$) for synthesis temperatures 700 °C and 900 °C, respectively, (**Figure 2.b.II**) is attributed to the increase in edge defects in NT-3DFG that allow D peak activation.[15] Upon further increase in NT-3DFG synthesis temperature, $I_D$ $I_G^{-1}$ declined; The decline was rationalized by high defect densities that limit the available number of near-by aromatic carbon rings that are necessary for D peak activation.[24, 25] For NT-3DFG, it is hypothesized the out-of-plane graphene edges brought in close proximity screen photon absorption in the highly D peak-active regions of the nearby intact $sp^2$ lattices. For extended NT-3DFG synthesis times (60 and 120 min) at 1100 °C, the $FWHM_G$ saturates at ~60 $cm^{-1}$ (**Figure 2.e.I**), and $I_D$ $I_G^{-1}$ decreases and saturates at ~2.4 (**Figure 2.e.II**), suggesting that highest edge defect densities for NT-3DFG have been achieved at these conditions. Increases in $Disp_G$ and $FWHM_G$ are observed with an increase in bulk disorder in carbon structures.[23] In the case of bulk structural defects, a higher $I_D$ $I_G^{-1}$ corresponds to higher $Disp_G$ and $FWHM_G$, thus facilitating a qualitative marker for discrimination between disorder at the edges and in the bulk.[15, 21, 22] The lack of correlation between $I_D$ $I_G^{-1}$ and $Disp_G$ (**Figure 2.c,f** & **Table S2-6**) as well as $I_D$ $I_G^{-1}$ and $FWHM_G$ (**Figure S4**) suggests that significant contributions to the D peak arise from edge defects rather than bulk-structural defects.[21, 22, 26] We also note the similar Raman signatures of NT-3DFG and micro-Raman signatures at edges of vertically align carbon nanowalls,[27] which further supports 3DFG edges' contributions to the D peak.

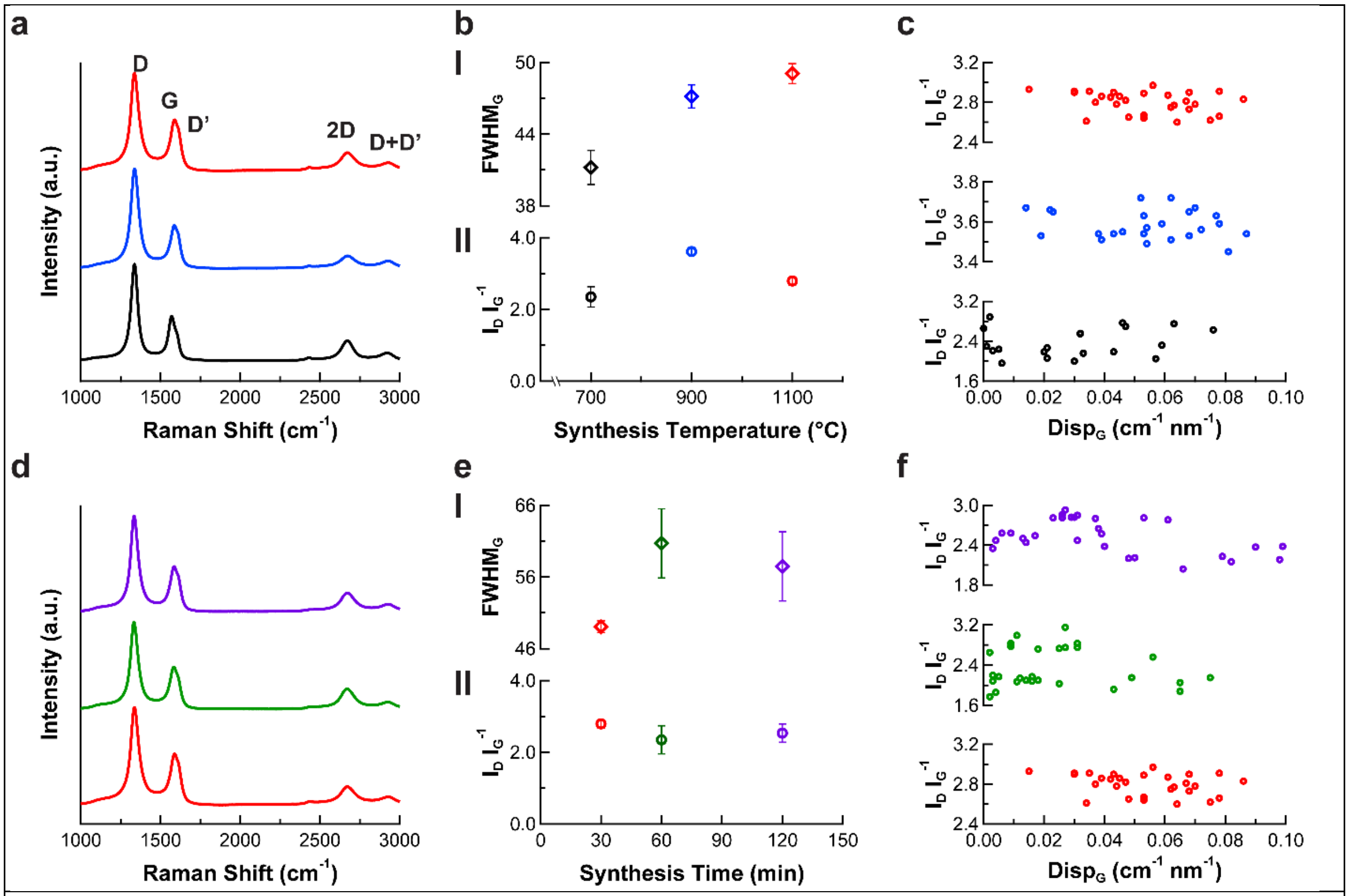

**Figure 2. Structural characterization of NT-3DFG. (a-c) Synthesis temperature effects (a)** Representative Raman spectra of NT-3DFG synthesized at 700 (**black**), 900 (**blue**), and 1100 °C (**red**) for 30 min. Peaks are annotated with labels: D, G, D' (shoulder to G peak), 2D, and D+D'. **(b.I)** Full width half maximum of the G peak ($FWHM_G$, **open diamonds**) and **(b.II)** D & G peak intensity ratio ($I_D$ $I_G^{-1}$, **open circles**) as a function of synthesis temperature. **(c)** $I_D$ $I_G^{-1}$ as a function of dispersion of the G peak ($Disp_G$.) **(d-f) Synthesis time effects (d)** Raman spectra of NT-3DFG synthesized at 1100 °C for 30 (**red**), 60 (**green**), and 120 min (**purple**). **(e.I)** $FWHM_G$ and **(e.II)** $I_D$ $I_G^{-1}$ as a function of synthesis time. **(f)** $I_D$ $I_G^{-1}$ as a function of $Disp_G$. Results are presented as mean ± SD ($n$ = 3).

To further clarify the nature of the graphene flakes in NT-3DFG, high-resolution transmission electron spectroscopy (HR-TEM) imaging was conducted (**Figure S5**). At low synthesis temperatures, multi-layered 3DFG flakes are visible at the circumferential surface of NT-3DFG, which appear similar in nature to folded turbostratic graphite.[28] At the 1100 °C synthesis temperature, many graphene flakes are visible at the circumferential surface which resemble suspended graphene sheets.[29]

**ORR activity of NT-3DFG.** Cyclic voltammetry (CV) experiments were conducted in 0.1 M KOH to assess the double-layer capacitance ($C_{dl}$), electrochemical surface area (ECSA), and the ORR response of NT-3DFG across all synthesis conditions. As reported in previous work,[15] increasing the size and density of 3DFG flakes results in significantly increased $C_{dl}$ and ECSA with as much as a ~ 43 times increase from NT-3DFG synthesized at 700 °C for 30 min to NT-3DFG synthesized at 1100 °C for 120 min (**Figure S6** & **Table S7**). All samples showed cathodic peaks under bubbled $O_2$ that were absent in the presence of pure $N_2$, indicating the reduction of $O_2$ at the NT-3DFG electrode (**Figure S7**). The formation of a second reduction peak occurred at lower potentials (~0.1 V vs. RHE). ORR proceeds through a 2 $e^-$ pathway or 4 $e^-$ pathway, and the presence of 2 peaks may indicate a two-step, 2 $e^-$ process to form $OH^-$ at high overpotentials.[30] Three synthesis conditions (700 °C for 30 min, 1100 °C 30 min, and 1100 °C 120 min) were chosen for rotating disk electrode (RDE) ORR testing, and samples were transferred to polished glassy carbon electrodes (GCE) (geometric surface area: 0.196 $cm^2$). The pronounced catalytic activity of NT-3DFG over GCE is demonstrated in representative linear sweep voltammetry (LSV) curves (**Figure 3.a**) for each synthesis condition performed at a scan rate of 10 mV $s^{-1}$ and rotational speed of 1600 rpm in $O_2$-saturated 0.1 M KOH. The catalysts achieved high limiting current densities (2.11 ± 0.17 mA $cm^{-2}$, 2.58 ± 0.20 mA $cm^{-2}$, and 2.95 ± 0.14 mA $cm^{-2}$) and exhibited excellent ORR onset potentials (0.739 ± 0.005 V, 0.766 ± 0.004 V, and 0.786 ± 0.009 V, measured as *E* versus RHE to achieve 0.1 mA $cm^2$ at 1600 rpm) for samples synthesized at 700 °C for 30 min, 1100 °C 30 min, and 1100 °C 120 min, respectively. We assess the current response within the kinetically-controlled regime using Tafel plots (**Figure 3.b** & **S8**), and the associated Tafel slopes (~ 55 mV $dec^{-1}$) suggest a pseudo two-electron pathway for all synthesis conditions.[31] To differentiate each catalyst's electrochemical performance and electronic conductivity, van der Pauw 4-point probe sheet resistance measurements were made (**Figure S9**).[32] We observed a significant decrease in NT-3DFG resistivity with increased density of 3DFG flakes[15, 33] (348 ± 30 mΩ cm, 171 ± 25 mΩ cm, and 69 ± 11 mΩ cm for samples synthesized at 700 °C for 30 min, 1100 °C 30 min, and 1100 °C 120 min, respectively). However, based on the low resistivities, negligible electronic *iR* losses across synthesis conditions (< 1 µV at a current of 0.5 mA and area of 0.196 $cm^2$) can be expected across the catalyst layer during RDE testing (**Table S8**).

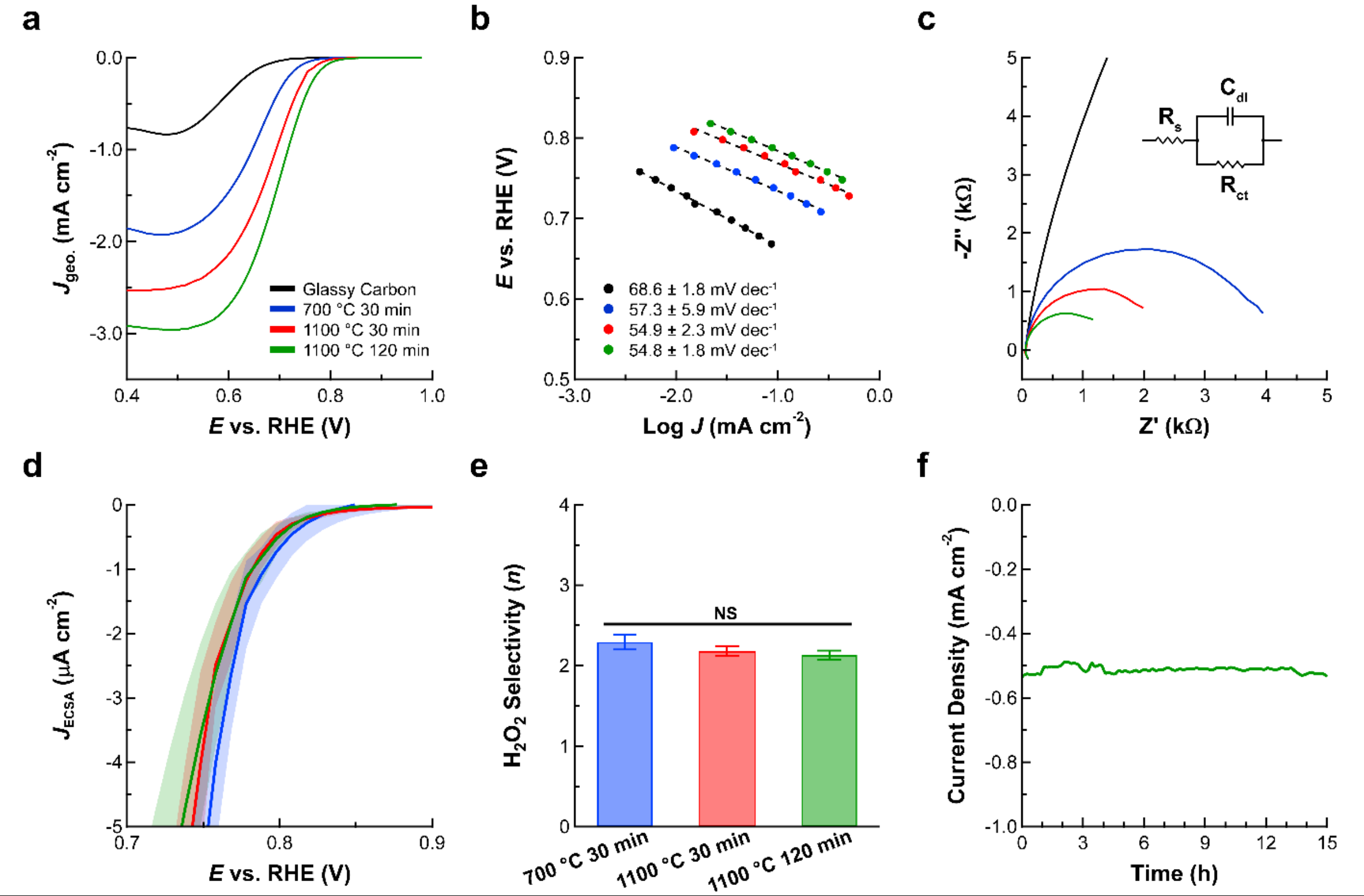

**Figure 3. Oxygen reduction reactions (ORR) with NT-3DFG. (a)** Representative rotating disk electrode linear sweep voltammetry (LSV) curves (*n* = 3) for polished glassy carbon (**black**) and NT-3DFG synthesized at 700 °C for 30 minutes (**blue**), 1100 °C for 30 min (**red**), and 1100 °C for 120 min (**green**) in $O_2$-saturated 0.1 M KOH at a scan rate of 10 mV $s^{-1}$ and rotational speed of 1600 rpm. **(b)** Tafel plots for LSV curves shown in **a** with linear fits for each slope denoted as a dashed line. **(c)** Representative electrical impedance spectroscopy (EIS) Nyquist plots (*n* = 3) for samples presented in **a**. Measurements were performed at 0.77 V vs. RHE. Inset displays circuit diagram used for modeling EIS response with $R_s$, $R_{ct}$, and $C_{dl}$ denoting solution resistance, charge transfer resistance, and double-layer capacitance, respectively. **(d)** LSV plot shows average ORR current normalized by electrochemical surface area ($J_{ECSA}$) with shaded area representing error ($J_{ECSA}$ ± SD). **(e)** $H_2O_2$ selectivity of NT-3DFG synthesized at 700 °C for 30 minutes (**blue**), 1100 °C for 30 min (**red**), and 1100 °C for 120 min (**green**). Electrodes were held at 0.5 V vs. RHE for 1 h and results are shown as mean ± SD (*n* = 3) of the ORR electron transfer number (*n*). NS indicates no statistical significance between conditions using ANOVA test and $p = 0.05$. **(f)** ORR current from NT-3DFG held at 0.5 V vs. RHE for 15 hours.

We further confirm the ORR activity of NT-3DFG through electrical impedance spectroscopy (EIS) measurements. Representative Nyquist plots from EIS spectra (taken at 0.77 V versus RHE in $O_2$-saturated 0.1 M KOH) and circuit model are shown in **Figure 3.c**. The semicircular Nyquist plots indicate parallel capacitive ($C_{dl}$) and resistive (ORR charge transfer resistance, $R_{ct}$) components with smaller semicircles indicating lower $R_{ct}$.[34] The measured $R_{ct}$ for NT-3DFG catalysts (**Table S9**) are consistent with recorded

ORR activity trends, with NT-3DFG synthesized at 1100 °C for 120 min reducing $R_{ct}$ by a factor 35 with respect to glassy carbon. We attribute the enhanced catalytic activity across NT-3DFG synthesis conditions to the increased number of ORR active sites as 3DFG flake density (correlated with ECSA) increases. **Figure 3.d** displays ECSA-normalized ORR activity for NT-3DFG catalysts shown as the average activity for each condition with a shaded error region (mean ± SD, *n* = 3). Comparable ECSA-normalized ORR activities as well as Tafel slopes across tested NT-3DFG synthesis conditions indicate the presence of similar type(s) of active site with comparable ORR kinetics, and that the active site density increases with increasing density of 3DFG flakes.

To assess the $H_2O_2$ selectivity of NT-3DFG, Koutecky-Levich analysis was performed (**Figure S10**), and the calculated ORR transfer numbers (*n*) were ~ 2.5 to 3 for tested catalysts. While the KL analysis suggests a preference towards a 2-electron pathway, the absolute values of *n* may be higher than expected due to the porosity of the catalyst layer and assumptions in KL theory (e.g. elementary reactions and constant *n* with changing angular velocity).[35] To directly measure $H_2O_2$ selectivity, ORR reaction products were quantified by chronoamperometry (CA) (**Figure S11**) and iodometric titration of $H_2O_2$.[36, 37] The $H_2O_2$ selectivity of NT-3DFG (measured as the ratio of the charge transferred during ORR at 0.5 V vs. RHE to the number of $H_2O_2$ molecules measured via titration) was 2.29 ± 0.09, 2.18 ± 0.06, and 2.13 ± 0.06 for samples synthesized at 700 °C for 30 min, 1100 °C for 30 min, and 1100 °C for 120 min, respectively (**Figure 3.e** & **Table S10**). NT-3DFG synthesized at 1100 °C for 120 min exhibited a $H_2O_2$ selectivity of 93 ± 3 % which is higher or comparable to precious metal alloy catalysts[4] and other nanocarbon catalysts.[6-8] The stability of NT-3DFG electrodes during ORR was investigated by 15-hour CA testing (**Figure 3.f**), and negligible change in the current output was observed. The NT-3DFG electrodes remained intact after ORR testing as evidenced by SEM (**Figure S12**) and Raman analysis (**Figure S13** & **Table S11**). NT-3DFG's low overpotential for ORR, high reaction selectivity, and electrochemical stability demonstrate ideal catalytic behavior for $H_2O_2$ generation under alkaline conditions.

**Surface state of NT-3DFG. Figure 4.a** provides a representative X-ray photoemission spectroscopy (XPS) survey scan for NT-3DFG before and after ORR (CV and $H_2O_2$ selectivity testing). Elemental surveys measured 11.5 ± 0.5 at. % oxygen which indicated significant oxygen incorporation at the surface of NT-3DFG after ORR (**Table S12**). The primary C1s feature at 284.1 ± 0.2 eV binding energy (BE) was attributed to $sp^2$ carbon, and the broad peak at 290.4 ± 0.2 eV BE was assigned to the **π–π*** plasmon loss feature (**Figure 4.b**).[38] After ORR testing, oxygen-carbon peaks were present in the C1s spectra (**Figure 4.c**) which were prescribed to C–O, C=O, and O=C–O bonds at binding energies of 285.7 ± 0.4, 286.8 ± 0.3, and 288.1 ± 0.4 eV, respectively (**Table S13**).[39, 40] The O1s peak (**Figure 4.d**) was fit with two peaks for C–O and C=O bonds at binding energies of 532.9 ± 0.1 and 531.6 ± 0.2, respectively. The ratios of the oxygen groups in the O1s peak and C1s peak analysis suggest the hydroxyl groups followed by the carbonyl groups were the most prominent oxygen species on the surface after ORR.

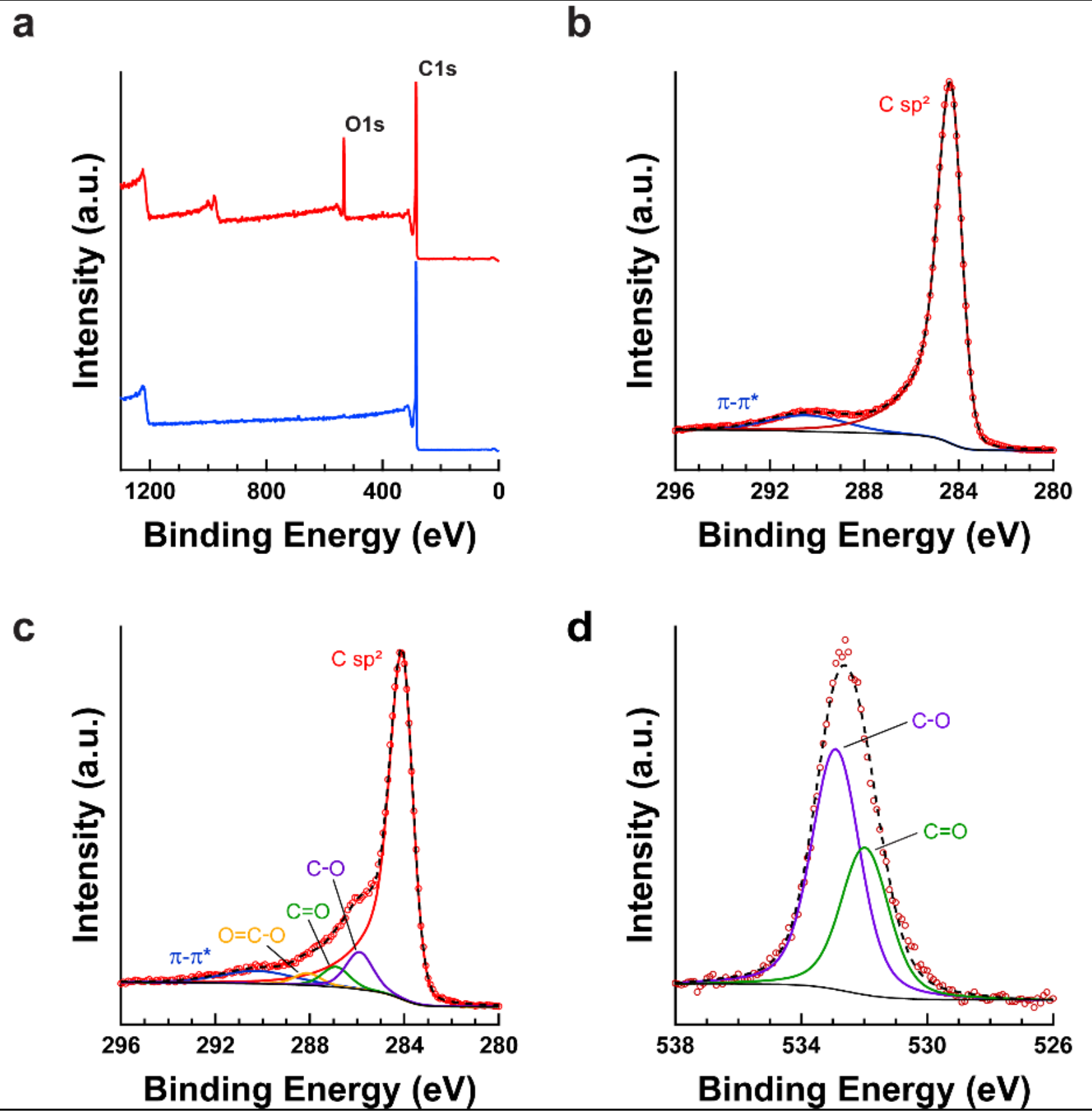

**Figure 4. Surface composition of NT-3DFG before and after ORR. (a)** Representative XPS survey scans ($n$ = 3) of NT-3DFG before ORR testing (**blue**) and NT-3DFG after ORR testing (**red**). **(b, c)** C1s chemical state analysis for NT-3DFG before and after ORR testing, respectively. **(d)** O1s chemical state analysis for NT-3DFG after ORR testing. NT-3DFG was synthesized at 1100 °C for 120 minutes. ORR testing included CV and $H_2O_2$ selectivity.

Electron energy loss spectroscopy (EELS) was conducted on NT-3DFG after ORR testing, and a prominent oxygen $K$ edge signature was observed in spectra taken from sample edges (**Figure 5.a,b**). In spectra collected 50 nm away from the edge and at pristine NT-3DFG edges, no oxygen signature was detected. The lack of a graphitic signature in the extended fine structure at the edges before and after ORR was further indication that single- to few-layer graphene was present at the edges[15, 33, 41] and stable after 10 h of ORR testing. EELS analysis of the near-edge carbon *K* fine structure for samples after ORR exhibited additional states (**Figure 5.c**) between 286-289 eV following the 1s to **π*** transition at ~ 285 eV (pristine graphene).[41]

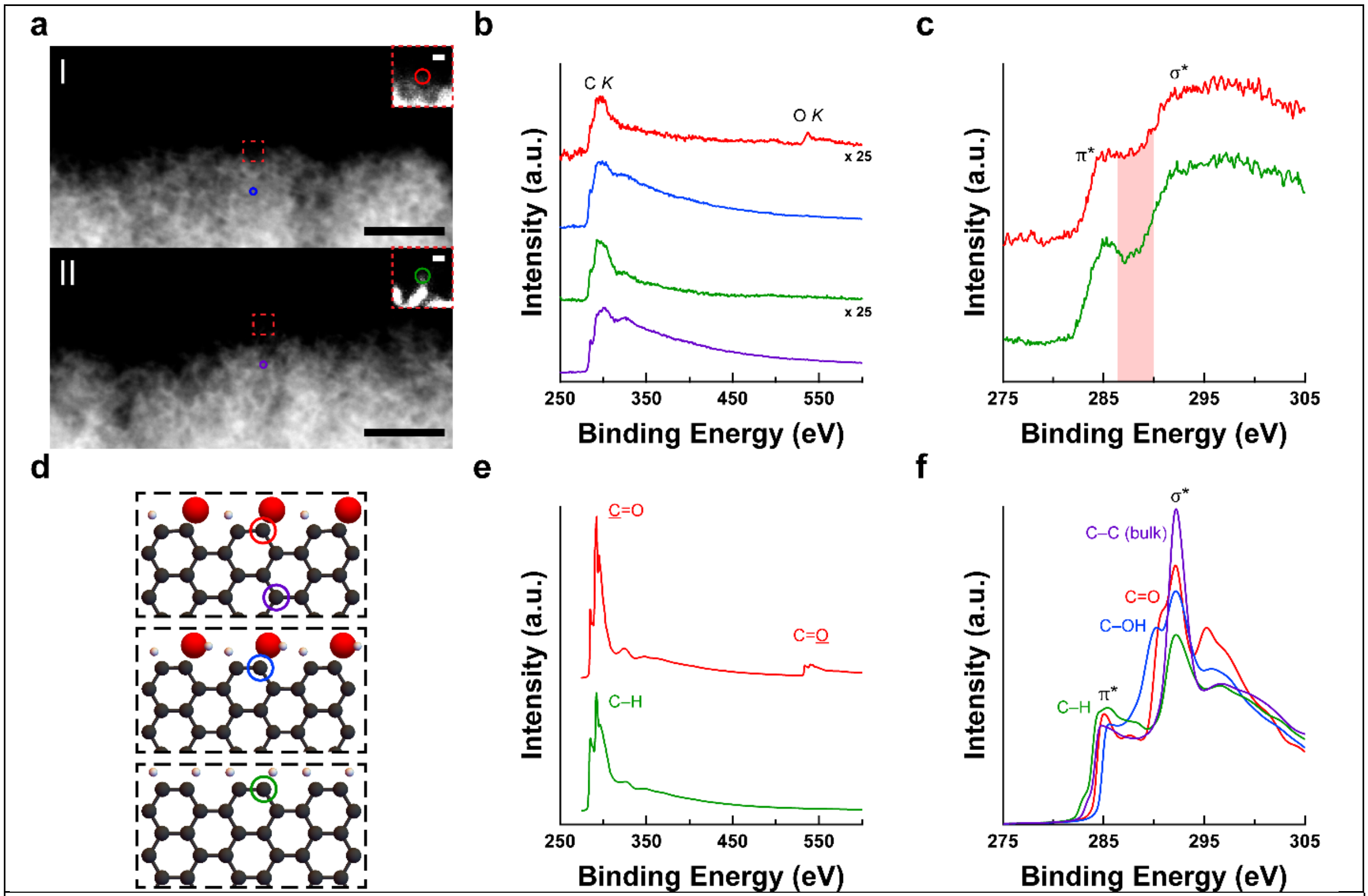

**Figure 5. Edge state of NT-3DFG before and after ORR testing. (a)** STEM images taken at the edge of NT-3DFG samples after **(I)** and before **(II)** ORR testing. Insets show edges of graphene flakes. Scale bars: 100 nm. Inset scale bars: 5 nm. **(b)** Representative EELS spectra (*n* = 3) taken from edge of pristine (**green**) NT-3DFG sample and sample after ORR testing (**red**). EELS spectra taken 50 nm away from edge for pristine (**purple**) sample and sample after ORR (**blue**). Circles with corresponding spectra color indicate location on samples where spectra were collected. **(c)** Carbon K edge EELS spectra taken from edge of pristine (**green**) NT-3DFG sample and sample after ORR testing (**red**). Filled region between 285 and 290 eV corresponds to additional states due to ORR testing. **(d)** Armchair edge terminations considered for ELNES calculations. O, H and C atoms are represented using red, silver and black spheres. **(e)** Theoretical EELS spectra associated with the C–H group (**green**) and C=O group (**red**), which are in agreement with the experiment (panel b). **(f)** Theoretical ELNES spectra for armchair edge configurations with spectra colors corresponding to respective highlighted C atoms (panel d).

We performed energy-loss near-edge fine structure (ELNES) simulations to understand the contributions of different functional groups at edge sites (**Figures 5.d** and **S14.a**). The near edge fine structure spectra is consistent with experiments and prior work exhibiting the 1s to **σ*** transition peak at 292.2 eV and the 1s to **π*** transition peak at 285.2 eV (**Figures 5.f and S14.c**).[42] Both theoretical spectra and experiments show distinct oxygen *K* edge contributions in the presence of C=O and C–O groups (**Figure 5.e** and **S14.b**), which is absent for pristine (C–H) edge configurations. The observed additional states (**Figure 5.c**) in the carbon *K* edge fine structure (**Figure 5.f**) were attributed to C–O and C=O bonds which is in agreement with XPS analysis for NT-3DFG and EELS spectra for graphene oxide.[42] Furthermore, in agreement with reported trends,[42] we find that the pre-edge peak associated with the carbonyl group corresponds to a higher binding energy relative to the hydroxyl. We note a reduction in the carbon *K* edge intensity for the **σ*** peak at functionalized edges relative to bulk C atoms, which can be attributed to the influence of termination groups on the $sp^2$-hybridized C orbitals. We observe negligible intensity changes in the **π*** peak ($p_z$ C orbital) after edge site functionalization, which primarily influences the in-plane $sp^2$-hybridized orbitals.

We draw distinctions between zigzag and armchair edge sites by comparing the near-edge carbon *K* fine structure for various terminations (**Figure 5.f** and **S14.c**). We find that the pre-edge peak for C–O and C=O bonds shift to higher energies for ZZ edges, which is consistent with carbon atoms at ZZ edges strongly binding oxygen relative to AC edges (1.251 Å and 1.234 Å carbonyl bond lengths respectively). This is in agreement with the Pourbaix diagram (**Figure 6.b**) where we observe the stability of high coverage carbonyl groups at ZZ edge sites. In addition, the overlap of the ELNES C=O peak with the **σ*** peak is due to stronger bonding at ZZ edges resulting in overall **σ*** peak broadening around 291.2 eV (**Figure S14.c**).

**DFT Calculations.** XPS and EELS analysis revealed that NT-3DFG edge sites saturate with oxygen-containing functional groups, which could significantly change the local electronic structure of active sites and thereby the ORR activity. We employed DFT calculations to identify the nature of active sites, which was central to understanding the observed activity and pathway selectivity. The pristine state of active sites is modeled on a two-dimensional graphene sheet for both the ZZ and AC edge terminations (**Figure 6.a**). On all surface sites, we determine the activity for hydrogen peroxide production through the adsorption energetics of the OOH intermediate by considering the associative mechanism.[4-6] While assessing the free energy landscape for hydrogen peroxide formation, it is worth noting that we consider the dissociated form ($OOH^-$) of $H_2O_2$ (pKa=11.7) under operating conditions (pH=13). While identifying potential active sites on the pristine surface can provide valuable information,[6, 9] the state of the surface under reaction conditions

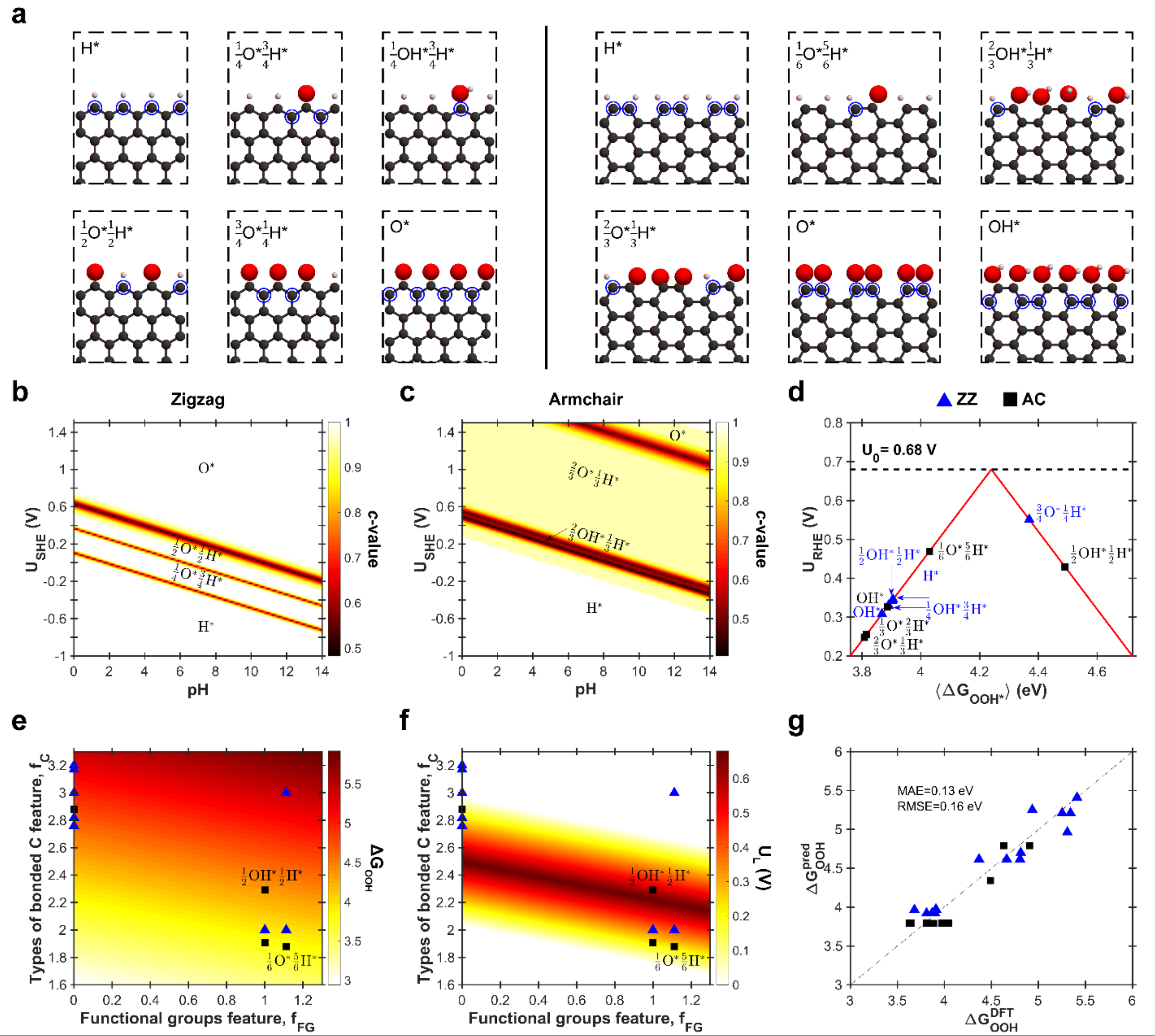

**Figure 6. DFT modeling for active site determination and inverse design through a geometric activity descriptor. (a)** Representative set of termination configurations of zigzag (ZZ) and armchair (AC) edge configurations with circled atoms showing the determined most stable active site for hydrogen peroxide formation. **(b, c)** Pourbaix diagrams showing the most energetically favorable surface species configurations for ZZ and AC edge configurations respectively, as a function of reaction conditions of electrode potential and pH. The color bar indicates the associated confidence value (c-value) with the predicted surface state, which quantifies the degree of agreement between exchange correlation functionals at the chosen level of DFT complexity. **(d)** Theoretical activity volcano for 2-electron oxygen reduction showing the activity predictions of various surface states. **(e)** Free energy of the OOH* intermediate, $\Delta G_{OOH}$, as a function of the local (nearest neighbor) coordination environment around the carbon active site, calculated from the proposed geometric free energy descriptor that unifies ZZ and AC edge sites. $f_{FG} = 0.40N_{OH} + 0.45N_{H}$, and is a feature of the catalytic site that captures the effect of functional groups. $f_C = 1.44N_{C-C} + 1.27N_{C-O} + 1.31N_{C-OH} + 1.86N_{C-H}$, and is a feature that incorporates the effects of bonded C types. **(f)** Predicted activity (limiting potential) towards hydrogen peroxide formation as a function of the local environment of possible active sites. **(g)** Parity plot showing the effectiveness of the constructed geometric descriptor for activity over a wide ($\approx$ 2.5 eV) descriptor range.

must also be considered; the importance of integrating surface Pourbaix diagrams in determining electrocatalytic activity and selectivity has been demonstrated elsewhere.[43-47] In this work, we explicitly perform a self-consistent loop between identifying the most stable surface state and determining the activity (active site) by constructing surface Pourbaix diagrams on basal and edge sites. The most energetically favorable surface states of active sites as a function of pH and electrode potential are determined from several explored species coverage configurations of oxygen functional groups such as hydroxyl (C–OH) and carbonyl (C=O) groups. Consistent with prior studies, our thermodynamic analysis identifies basal sites to be inactive towards hydrogen peroxide formation.[6] The analysis suggests that both AC and ZZ edge sites including those with hydrogen defects are saturated by carbonyl groups with coverage $\geq 0.5$ monolayer (ML) under reaction conditions of pH = 13 and limiting potentials relevant (≈0.7 V vs. RHE) for hydrogen peroxide formation. The stable surface states are shown as a function of pH and $U_{SHE}$ for ZZ and AC edge sites in the corresponding probabilistic Pourbaix diagrams (**Figures 6.b** and **6.c**, respectively) in which each predicted stable surface configuration is associated with a confidence value (c-value ∈ [0,1]) to quantify the degree of agreement at the chosen level of DFT fidelity. This approach provides a way to address a major challenge inherent to DFT, which is the sensitivity of determined surface phases towards the choice of exchange-correlation functional,[48, 49] by leveraging a recently developed method that uses Bayesian error estimation techniques to compute the uncertainty associated with our predictions.[50] Based on this thermodynamic analysis, we find that AC sites with low ($\frac{1}{6}$ML) coverage and ZZ sites with high ($\frac{3}{4}$ML) coverage of carbonyl groups exhibit high (overpotential $< 0.2$ eV) activity (**Figure 6.d**). Although the active sites on the stronger binding leg of the volcano exhibit higher activity towards oxygen reduction to water from a thermodynamic standpoint,[51, 52] favorable kinetics towards selective hydrogen peroxide formation is expected on the basis of agreement with experiments.[53, 54] We rationalize the trends in surface energetics between various active sites, for example the stronger OOH*-binding nature of pristine (H-covered) AC edge sites relative to ZZ edge sites, based on electronegativity differences and associated Bader charges between the active site carbon atoms. In order to further understand and capture the effects of various bonding interactions around the active site on the catalytic activity, a geometric descriptor for the activity towards hydrogen peroxide is derived on the basis of the active site coordination environment, which enables activity predictions of graphene surface sites with high accuracy. We arrive at the model on the basis of maximizing the Bayesian information criterion given by $\Delta G_{OOH^*}^{pred}(eV) = 0.63 + 0.40N_{OH} + 0.45N_{H} + 1.44N_{C-C} + 1.27N_{C-O} + 1.31N_{C-OH} + 1.86N_{C-H}$ with a mean absolute error (MAE) of ≈0.1 eV (**Figure S15** &

**Table S14**), where $N_{OH}$ and $N_H$ represent the presence of OH and H functional groups adjacent to the active carbon site of interest respectively. $N_{C-C}$, $N_{C-O}$, $N_{C-OH}$ and $N_{C-H}$ represent the counts of adjacent $sp^2$ hybridized carbon atoms with all (three) carbon neighbors, all carbon neighbors except one -O, all carbon neighbors except one -OH, and all carbon neighbors except one -H, respectively (**Figure 6.f**). The relative magnitudes of coefficients in the model emerge to be in agreement with what is expected in terms of the weakening of OOH*-binding caused by the presence of each kind of moiety adjacent to the active site carbon, for instance the largest weakening is expected to result from an adjacent carbon with all carbon neighbors. The geometric descriptor unifies AC and ZZ edge configurations and allows us to predict the landscape of surface energetics on graphene surfaces reasonably accurately, in addition to enabling inversion of the design problem to identify the nature of optimal active sites to be engineered for desired adsorption characteristics. Our analysis reveals that highly active sites exhibit optimal local coordination of one or two adjacent carbons that are saturated by oxygen terminations, which provides directions for rational material design within a broad range of carbon-based two-dimensional materials and heterostructures.

## Discussion

NT-3DFG achieves promising two-electron ORR activity and selectivity without additional pretreatment steps such as surface modification under concentrated acids. We attribute this performance to NT-3DFG's hierarchical structure that allows for many exposed single-layer graphene edges that readily become oxygenated *in situ* under alkaline ORR conditions where O content has been shown to enhance ORR activity.[6, 8] We observe a similar ECSA-normalized ORR activity for samples synthesized at 700 °C and 1100 °C which may suggest: the PECVD growth method results in similar types or distributions of edge terminations across the tested synthesis temperature space or the pristine state of the edges does not play a significant role in determining ORR activity.

Our experimental and computational findings suggest that NT-3DFG is composed of both functionalized AC and ZZ edges during and after ORR. The presence of both edge types is consistent with proposed diffusion-limited aggregation growth models for NT-3DFG that indicate the formation of many possible edge states, which is further supported by small differences in computed formation energies.[55] We posit a significant presence of armchair edges in NT-3DFG based on the hydroxyl XPS signature (**Figure 5.c**) (relative to carbonyl) together with the predicted stability of OH* termination ($\frac{2}{3}$OH*$\frac{1}{3}$H*) only for AC edges (**Figures 6.b,c**). In addition, both ZZ and AC (with lower confidence) edges are predicted to exhibit O* termination ($\frac{1}{2}$O*$\frac{1}{2}$H* and $\frac{2}{3}$O*$\frac{1}{3}$H*, respectively), suggesting that the reported carbonyl XPS signature emerges from a combination of edge site types. We corroborate our XPS peak assignments to graphene edges through localized EELS spectra and ELNES, which are consistent with predicted edge terminations from surface Pourbaix diagrams. In addition, the increase in catalytic activity with increasing density of 3DFG flakes offers further confirmation that the identified ORR active sites correspond to out-of-plane graphene flake edges. We report low computed overpotentials for ORR (within $\approx 0.1$ V) for both AC and ZZ active sites ($\frac{3}{4}$O*$\frac{1}{4}$H* ZZ, $\frac{1}{6}$O*$\frac{5}{6}$H* AC, and $\frac{1}{2}$OH*$\frac{1}{2}$H* AC configurations), and these oxygen-modified site configurations show similar coverages to the predicted stable phases from the Pourbaix diagram. Zigzag edges with high coverage of carbonyl groups can explain XPS observations as well as predicted thermodynamic stability, although explicit demonstration of their presence and contribution to ORR activity is a current challenge.

Our experiments and theoretical analysis suggest engineering ORR active sites of edge-oriented graphene electrocatalysts through electrochemical pretreatment to generate specific edge configurations. We point out the possible kinetic locking of highly active sites that may not be thermodynamically stable from the Pourbaix diagram, such as the $\frac{1}{6}$O*$\frac{5}{6}$H* AC edge configuration. While the observed ORR activity of NT-3DFG may result from kinetically stabilized edge site configurations that are predicted to exhibit high activity, further kinetic studies are necessary to provide more insight. We also note that the incorporation of other potential edge configurations (e.g., mixed edges, edges at plane corners, and step edges) may further alter ORR activity, although targeted synthesis and experimental confirmation of specific edge configurations will require further study.

In summary, we report a novel graphene-based nanomaterial and controlled synthesis procedure that aids in understanding the origin of high selectivity and activity for two-electron ORR in $sp^2$-hybridized carbon structures. The significance of graphene edges sites towards two-electron ORR is demonstrated by the correlation of NT-3DFG edge density and ORR activity. NT-3DFG exhibits superior performance with hydrogen peroxide selectivity of 93 ± 3 % and onset potential of ca. 0.79 V. Our combined experiment-theory approach suggests that edge sites are saturated by carbonyl (C=O) or hydroxyl (C–OH) groups under reaction conditions. We predict zigzag edge sites with high carbonyl group coverage as well as other edge configurations with a similar local coordination environment may enable selective, two-electron ORR. Our findings and proposed model provide an increased understanding of two-electron ORR on graphene edge-based structures and offer directions for future carbon-based electrocatalyst design.

## Methods

**Nanowire Template Synthesis.** Gold nanoparticle (AuNP)-catalyzed vapor-liquid-solid (VLS) growth was used to synthesize Silicon nanowires (SiNWs), following previously described synthesis.[15, 56] In short, a 1.5 cm × 2.0 cm (100) Si substrate with a 600 nm wet thermal oxide layer (Nova Electronic Materials) or a 1.5 cm × 1.5 cm fused silica substrate (University Wafer, used for electrochemical measurements) was cleaned with acetone in an ultrasonic bath for 5 min, washed with IPA, and $N_2$ blow-dried. The substrate was placed in a UV−ozone system (PSD Pro series digital UV-Ozone, Novascan) for 10 min at 150 °C. The substrate was then functionalized with 450 μL (400 μL for 1.5 cm × 1.5 cm substrate) of 4:1 deionized (DI) water/poly-L-lysine (0.1% w/v, Sigma-Aldrich, Catalog No. P8920) for 8 min. Following this step, the substrate was gently washed three times in DI water and gently $N_2$ blow-dried. 450 μL (400 μL for 1.5 cm × 1.5 cm substrate) of 4:1 DI water/30 nm AuNP solution (Ted Pella, Inc., Catalog No. 15706-1) was dispersed onto the PLL-coated substrate for 8 min. The substrate was gently washed three times in DI water, gently $N_2$ blow-dried, and introduced into a custom-built CVD setup. After a baseline pressure of $1*10^{-5}$ Torr was attained, the temperature was ramped up to 450 °C in 8 min and the system was allowed to stabilize at 450 °C for 5 min. Nucleation was then conducted at 450 °C for 15 min under 80 standard cubic centimeters per minute (sccm) $H_2$ (Matheson Gas) and 20 sccm $SiH_4$ (10% in $H_2$, Matheson Gas) at 40 Torr. SiNW growth was then conducted at 450 °C for 100 min under 60 sccm $H_2$, 20 sccm $SiH_4$, and 20 sccm $PH_3$ (1000 ppm in $H_2$, Matheson Gas) at 40 Torr. The sample was then cooled to room temperature at base pressure.

**Mesh Formation.** To form SiNWs mesh, the SiNWs were collapsed by flowing liquid $N_2$ into the CVD quartz tube at room temperature under 200 sccm Ar flow. Then the system was evacuated to base pressure and the SiNWs mesh was annealed at 800 °C for 10 min under 200 sccm $H_2$ flow at 1.6 Torr. The sample was then cooled to room temperature at baseline pressure.

**Out-of-Plane 3D Fuzzy Graphene Synthesis.** A PECVD process was used to synthesize 3DFG on SiNWs mesh template. A SiNWs mesh was introduced into a custom-built PECVD system. The sample was centered in the tube on a carrier wafer that was 4.0 cm from the edge of the RF coil and outside the furnace. The furnace temperature was ramped up to the synthesis temperature of 700, 900, and 1100 °C in 13, 18, and 25 min, respectively, under 100 sccm Ar flow at 0.5 Torr. The system was then stabilized for 5 min under 100 sccm Ar (Matheson Gas) flow at 0.5 Torr. 3DFG synthesis was carried out at the synthesis temperature for varying synthesis time periods (30, 60, and 120 min) under 25 mTorr of $CH_4$ (5% $CH_4$ in Ar, Airgas) at 0.5 Torr. Inductively coupled plasma was generated 10 seconds before the start of the synthesis step using a 13.56 MHz RF power supply (AG 0313 Generator and AIT-600 RF, power supply and auto tuner, respectively, T&C Power Conversion, Inc.). The plasma power was held at 50 W. The furnace was moved over the sample following plasma ignition. The plasma was shut off after each synthesis step and the sample was rapidly cooled to 100 °C under 100 sccm Ar before removal from the CVD system. Three independently synthesized samples were synthesized for each PECVD synthesis condition.

**Scanning Electron Microscopy (SEM) Imaging.** SEM imaging was conducted using a FEI Quanta 600 field emission gun SEM. Micrographs were acquired at 20 kV accelerating voltage with a 5 mm working distance. NT-3DFG diameter was measured from high-resolution (2048 × 1768 pixels) SEM micrographs in ImageJ for 90 nanowires, across three independently synthesized samples, for each synthesis condition. NT-3DFG diameter is presented in **Figure 1e** as the mean ± SD ($n$ = 90) per synthesis condition. The 3DFG growth rate, $k$, was approximated by the change in radius of the nanowire before and after synthesis with the following equation:

$$\mathrm{k} = \frac{1}{2}\left(\frac{\mathrm{d_{NT-3DFG}} - \mathrm{d_{SiNW}}}{\mathrm{t}}\right)$$

where $d_{NT-3DFG}$ is the NT-3DFG diameter, $d_{SiNW}$ is the diameter of the nanowire template, and $t$ is the synthesis time. The 3DFG growth rates across synthesis conditions were used to calculate the activation energy for 3DFG formation using the Arrhenius equation[57]:

$$\mathrm{k} = \mathrm{k_o} \times \exp\left(-\frac{\mathrm{E_A}}{\mathrm{RT}}\right)$$

where the activation energy ($E_A$) for synthesis represents the minimum energy required for growth of the out-of-plane graphene flake structures of NT-3DFG. Edge density analysis (**Figure S16**) was performed to determine NT-3DFG flake edges along individual nanowires using Matlab and a Canny edge detection protocol with least count and edge connections.[58] Using an Au/C calibration specimen (High Resolution Gold on Carbon, TedPella, Catalog No. 617) and measuring the pitch of 30 linescans across Au nanoparticles, the SEM resolution was determined to be 5 nm. Edge features below the resolution limit were excluded from edge analysis. Edge density measurements, presented as the mean ± SD (*n* = 30) per synthesis condition in **Figure 1f**, were performed on high-resolution SEM images for NT-3DFG synthesized at 700, 900, and 1100 °C for 30 min. The thicknesses of NT-3DFG meshes (**Table S8**) were measured by cleaving growth substrates after synthesis and imaging the samples in a 90° tilt SEM sample holder. Mesh thicknesses are presented as mean ± SD (*n* = 3).

**Diffusion-Limited Aggregation Based Growth Model Simulations.** The simulations performed are not a substitute for a full atomistic description that would consider time scale and local and global chemistry. However, since such simulation is currently out of reach of any modeling method, the proposed approach has several well-controlled features that enable us to shed light on the experimentally observed growth process for NT-3DFG. In the calculations performed, each nucleating *particle* consists of a single carbon dimer, and its movement is constrained to discrete locations on a hexagonal lattice. Notice that these rules explicitly constrain the structure to form honeycomb shapes in regions which are filled, and if one were to fill all sites in space, the resulting structure would be AA-stacked graphite. However, instead of modeling optimal packing the algorithm used explored the effect of diffusion-limited growth on the generation of large-scale nanostructures. To seed the process, we start with a fixed cylindrical core to mimic the SiNW used in the experimental conditions (**Figure S2**). The core is oriented along the direction of stacking in the lattice and is periodic in this direction. Dimers are introduced at the boundary of a large periodic cell centered at the core, and each step of Brownian motion moves the dimer to a randomly chosen neighboring lattice point, defined as the six nearest triangular lattice neighbors in the xy-plane and the two immediate neighbors along ±z. Lattice parameters do not enter explicitly into the formalism; the particle's destination each step is chosen randomly and uniformly from those neighbors which are vacant. This means that movement may be anisotropic when interpreted spatially, as there is generally a 75% chance of movement within the xy-plane versus a 25% chance of movement along z, regardless of the relative physical lengths of these motions. Nucleation requires a point with two occupied adjacent sites in the xy-plane which will form a complete triangle with the new dimer. This rule specifically promotes the creation of layers, as nucleation in the z-direction is thus impossible. Together with the hexagonal lattice structure, this strongly promotes local $sp^2$ chemistry. In this simplified model, the effect of temperature is accounted for with the introduction of an integer parameter $n_{touch}$. During an event, the code tracks how many times the dimer has touched each eligible nucleation point. As soon as a point has been visited $n_{touch}$ times, the nucleation event ends. An increased $n_{touch}$ corresponds to a greater temperature, as it represents the increased kinetic energy of the dimer. We note that while greater values of $n_{touch}$ are expected to correspond to greater temperatures, an explicit relationship is not known. $n_{touch}$ is a coarse parameter; and a finer-grained control could be achieved by replacement with a probability parameter, $p_{end}$, of the event ending when an eligible point is touched. Varying synthesis temperature likely alters the time scale between nucleation events (see **Figure S1**), and therefore we do not expect simulations to represent equal intervals of time. Results shown correspond to 600,000 nucleation events with a NW diameter of 30 nm and unit cell length of 27 nm.

**Transmission Electron Microscopy (TEM) Imaging.** TEM imaging was conducted using a FEI Titan G2 80−300 Cs-corrected TEM/STEM fitted with a high resolution Gatan Imaging Filter Tridiem energy-filter. High-resolution TEM (HRTEM) images were acquired at 300 kV accelerating voltage with 2048 × 2048 slow-scan CCD cameras.

**Electron Energy Loss Spectroscopy (EELS).** EELS spectra were collected with FEI Titan G2 80−300 Cs-corrected TEM/STEM in dual STEM/EELS mode. Spectral scans were taken as integrated counts over a 20 s dwell time at dispersions of 0.3 and 0.1 eV per channel. Line scans were performed from the exterior to the interior of the sample at 5 nm increments to locate the edge of the sample. Spectra were background subtracted using FEI ES Vision software. NT-3DFG samples were synthesized at 1100 °C for 30 min. ORR testing for NT-3DFG samples used for EELS analysis consisted of ORR CV testing and electrochemical stability testing for 10 h.

**Raman Spectroscopy.** Raman spectroscopy was conducted using a NT-MDT NTEGRA Spectra (100x objective) with 532 nm and 633 nm excitation wavelengths at a laser power of 2.38 mW for both wavelengths. Dispersion of G peak ($Disp_G$) calculations were made from spectra from the same point with the 2 excitation wavelengths. $Disp_G$ is calculated as:

$$\mathrm{Disp_G} = \left| \frac{\mathrm{Pos(G)_{633\ nm}} - \mathrm{Pos(G)_{532\ nm}}}{\lambda_{633\ nm} - \lambda_{532\ nm}} \right|$$

Where Pos(G) is the spectral position of the G peak under a given excitation laser and λ is the excitation laser's wavelength. Raman spectra were acquired from 10 randomly distributed points across each NT-3DFG sample with three independently synthesized samples per synthesis condition. Correlation tests were conducted for $Disp_G$ & $I_D\ I_G^{-1}$ as well as $FWHM_G$ & $I_D\ I_G^{-1}$ for all synthesis conditions with the equation:[59]

$$r = \frac{\sum(x_i - \bar{x})(y_i - \bar{y})}{\sqrt{\sum(x_i - \bar{x})^2 (y_i - \bar{y})^2}}$$

where r is the coefficient of linear correlation, x is the $Disp_G$ and y is the respective $I_D\ I_G^{-1}$.

**X-ray Photoemission Spectroscopy (XPS).** XPS was conducted using an Escalab 250Xi (Thermo Scientific) with Al Kα (1486.68 eV) x-ray source at ultra-high vacuum ($10^{-9}$ Torr). Survey spectra were acquired by averaging 5 scans with a 120-eV pass energy, 15 ms acquisition time, and 0.5-eV step size. O1s and C1s spectra were acquired by averaging 10 scans with a 50-eV pass energy, 50 ms acquisition time, and 0.1-eV step size. Compositional and chemical state analyses were completed with Thermo Scientific Avantage peak fitting software as follows: Elemental peaks from survey scans were identified using Avantage peak auto-identification. Integrated peak areas in counts per second (CPS) were calculated using an integral (Shirley) background and a fitted Lorentzian-Gaussian (30 % L:G mix sum) function.[38] For chemical state analyses of the C1s peak, the asymmetric shape of the $sp^2$ peak was fit with a Doniac-Sunjic function that can be approximated by the convolution of the fitted peak with a power function of the type:

$$\frac{1}{E^{1-\alpha}}$$

where *E* is the binding energy and *α* is the singularity index. This function corresponds to the core-hole potential-screening spectrum of electron-hole pair excitations that occur at the Fermi level, and the function is commonly used for spectra of graphite and aromatic molecules.[38] The *α* for NT-3DFG was found to be 0.08 ± 0.02 which is agreement with literature values for graphitic materials.[38, 60] The broad plasmon shake up feature was fit with a FWHM of ~ 3 eV near ~ 290 eV binding energy (BE).[61] For fitting of the C1s peak after ORR testing, the fit parameters for the $sp^2$ peak (BE, FWHM, *α*) and plasmon peak (BE, FWHM) were fixed relative to pristine NT-3DFG. BE values from literature and the same FWHMs were used to fit oxygen functional groups.[39, 40] XPS was conducted for three independent NT-3DFG samples synthesized at 1100 °C for 120 min, and values were reported as the mean ± SD (*n* = 3).

**Sheet Resistance Measurements.** Sheet resistance (van der Pauw) measurements were performed as shown previously for NT-3DFG.[15] Briefly, flat Au-plated pogo pins were placed at the four corners of a 1 cm square pattern. A current sweep was performed from −100 to +100 μA, and potential change was recorded with a source meter unit (SMU, Keithley 2401). The NT-3DFG sample is assumed to be a flat surface with no pores (i.e., porosity of NT-3DFG samples was not taken into consideration). Sheet resistance measurements were performed with three independently synthesized samples per NT-3DFG synthesis condition and results are presented in **Figure S9** and **Table S8** as mean ± SD (*n* = 3).

**Electrochemical Cell Preparation.** Prior to conducting cyclic voltammetry testing, contacts composed of 5 nm Cr (99.99%, The R.D. Mathis Co.) followed by 150 nm Au (99.999%, Praxair) were evaporated at the edge of the NT-3DFG samples using a thermal evaporator (Angstrom Thermal Evaporator). Polystyrene wells were then sealed to the NT-3DFG samples using poly dimethyl-siloxane (PDMS) (Sylgard 184 Silicone Elastomer, Dow Corning). Geometric area of NT-3DFG exposed to the electrolyte after adhering the polystyrene well was determined by imaging the sample surfaces at 1× magnification with a stereo microscope (AmScope) and analyzing the images using ImageJ. In order to ensure the interaction of NT-3DFG with the electrolyte, IPA infiltration was conducted, as previously described.[15] 2 mL IPA was introduced into the well, and the sample was placed in a desiccator under vacuum for 5 min. 1.8 mL of IPA was aspirated out and 1.8 mL of DI water was introduced, followed by 5 min in the desiccator under vacuum. This process was repeated four times to exchange IPA with the DI water. A similar procedure was followed to exchange the DI water and 0.1 M KOH (Aqua Solutions, Catalog No. 7220) electrolyte.

**Cyclic Voltammetry (CV) testing.** CV was performed using a PalmSens3 potentiostat (Palmsens BV) in a three-electrode system with a NT-3DFG sample, Pt wire (CHI115, CH Instruments Inc.), and Ag/AgCl electrode in 3 M KCl (+0.209 V vs. Normal Hydrogen Electrode) (CHI111, CH Instruments Inc.) as the working, counter, and reference electrodes, respectively. Prior to each test, $N_2$ (99.9999%, Airgas) or $O_2$ (99.999%, Airgas) was bubbled through the cell for 15 min. CV measurements were recorded with a potential range of 0.2 to -1.2 V versus Ag/AgCl at 50 mV $s^{-1}$ $N_2$- or $O_2$-saturated 0.1 M KOH. CV current values for each sample were normalized with respect to the geometric area of the sample. For preliminary comparison of ORR activity of NT-3DFG samples across all synthesis conditions, ORR onset potential was measured at 0.1 mA $cm^{-2}$ after $N_2$-baseline current subtraction. Double-layer capacitance ($C_{dl}$) was determined by conducting CV with 0.1 M KOH electrolyte solution within a potential range from −1.2 to 0.2 V versus Ag/AgCl at increasing scan rates of 10, 50, 100, and 200 mV $s^{-1}$. Geometric current density at 0 V versus Ag/AgCl (where no Faradaic process took place) was plotted against the scan rate and the slope of the linear regression was computed to determine $C_{dl}$. The electrochemically active surface area (ECSA) was calculated (**Figure S7**) based on the measured $C_{dl}$ and an estimated specific capacitance 20 µF $cm^{-2}$.[62, 63] After the measurements, the sample was washed for 30 min in three DI water baths and dried overnight in a clean hood. CV testing was conducted for three independently synthesized samples for each NT-3DFG synthesis condition with five scans acquired for each CV test. Current densities and potentials are reported as the mean ± SD (*n* = 3).

**$H_2O_2$ Selectivity.** The $H_2O_2$ selectivity of NT-3DFG during oxygen reduction reaction (ORR) was determined by chronoamperometry (CA) and iodometric titration of $H_2O_2$[37] generated in the electrolyte solution. After CV testing, the sample was held at 0.5 V vs. RHE for 1 hour in $O_2$-saturated 0.1 M KOH. The current was recorded and integrated to calculate the total charged transferred. After CA, 1.0 ml of the electrolyte, along with 1.0 ml of 2 wt.% potassium iodide (KI) (≥99.0%, Sigma-Aldrich, Catalog No. 221945) solution, 1 ml of 3.5 M sulfuric acid ($H_2SO_4$) (96%, KMG Electronic Chemical Inc.), and 50 µL of a molybdate catalyst (1 g ammonium molybdate (99.98%, Sigma-Aldrich, Catalog No. 277908) in 10 mL of 6 N ammonium hydroxide (≥99.99%, Sigma-Aldrich, Catalog No. 338818), followed by adding 3 g ammonium nitrate (≥99.5%, Sigma-Aldrich, Catalog No. A7455) and diluting the solution to 50 mL with DI water) were added to a vial. $H_2O_2$ readily oxidized $I^-$ to $I_2$ under acidic conditions, and the presence of iodine was observed by a characteristic yellow-brown solution color. The solution was then titrated with a 10.000 ± 0.025 mL burette and dropwise additions of a 5 mM sodium thiosulfate ($Na_2S_2O_3$) (≥ 97.0%, Emplura, Catalog No. 106512) solution. After the solution reaches a straw color, ~5 mg of starch (ACS reagent, Sigma-Aldrich, Catalog No. S9765) is added as an indicator. Titration is complete when the purple color of the starch–iodine complex vanishes to yield a clear solution. The chemical reactions for iodometric titration of $H_2O_2$ are as follows:

$$\mathrm{H_2O_2 + 2KI + H_2SO_4 \rightarrow I_2 + K_2SO_4 + 2H_2O}$$

$$I_2 + 2Na_2S_2O_3 \rightarrow Na_2S_4O_6 + 2NaI$$

The volume of $Na_2S_2O_3$ titrated was measured, and the moles of iodine consumed during titration was calculated. One mole of iodine is formed for every mole of $H_2O_2$ present in the electrolyte solution. The concentration of $H_2O_2$ in the electrolyte is obtained, and the total amount of $H_2O_2$ generated is compared to the total charged transferred during CA. $H_2O_2$ selectivity testing was conducted for three independently synthesized samples, and selectivity is reported as the mean ± SD ($n$ = 3).

**Rotating disk electrode (RDE) mesh transfer.** A NT-3DFG mesh transfer method was developed to transfer NT-3DFG meshes from growth substrates to RDE. Prior to the NT-3DFG synthesis process, a 2 µm-thick $SiO_2$ layer was deposited on a (100) Si substrate using a Trion Orion PECVD System. NT-3DFG synthesis was performed as stated earlier in the methods section. After synthesis, the sample was placed in a 49% hydrofluoric acid bath for 90 min to etch the $SiO_2$ layer and lift off the NT-3DFG mesh. The sample was transferred to a DI water bath three times for 10 min each to remove residual HF. Prior to NT-3DFG transfer, the glassy carbon (GC) tip (area: 0.196 $cm^2$) (Pine Research, Catalog No. AFED050P040GC) of a RDE (Pine Research, Catalog No. AFE5TQ050PK) was polished with 0.05 µm alumina slurry (Allied High-Tech Inc., Catalog No. 90-187505) and a microcloth pad (Pine Research, Catalog No. AKPOLISH) until a smooth and reflective surface was achieved. The GC tip was washed three times with DI water then IPA. The NT-3DFG mesh was transferred onto the GC electrode tip and dried for 1 hour in a clean hood.

**Rotating disk electrode (RDE) ORR testing.** RDE ORR testing was conducted using a WaveVortex 10 rotator (Pine Research) and a PalmSens3 potentiostat in a three-electrode setup with NT-3DFG mesh on polished GC tip RDE, graphite electrode (Pine Research, Catalog No. AFCTR3B), and Ag/AgCl electrode in 4 M KCl (+0.199 V vs NHE) (Pine Research, Catalog No. RREF0021) as the working, counter, and reference electrode, respectively. $N_2$ was bubbled through 0.1 M KOH for 1 hour in a sealed cell. The RDE was slowly ramped up to a rotational speed of 1600 rpm and linear sweep voltammetry (LSV) measurements were made from 0.2 to -1.0 V versus Ag/AgCl at 10 mV $s^{-1}$. $O_2$ was then bubbled through the cell for 30 min, and EIS was performed following previous protocol at -0.2 V versus Ag/AgCl. RDE LSV measurements were repeated for the $O_2$-saturated cell. Tafel plots were generated with the Tafel equation:

$$E = C + b \times \log i_{\mathrm{k}}$$

where $E$, $C$, and $b$ denote the measured potential, exchange current constant, and Tafel slope, respectively. The kinetic current ($i_k$) was obtained by mass-transport correction with the following relation:

$$\frac{1}{i} = \frac{1}{i_\mathrm{k}} + \frac{1}{i_\mathrm{l}}$$

where $i$ is the recorded current and $i_l$ is the diffusion-limited current (maximum current value greater than 0.4 V versus RHE). Additional LSV scans were acquired at rotational speeds of 400, 800, 1200, 1600, and 2000 rpm for Koutecky-Levich analysis. The Koutecky-Levich equation relates electron transfer number to measured current and rotation speed as follows:

$$\frac{1}{i} = \frac{1}{i_\mathrm{k}} + \frac{1}{0.62nFAD_O^{\frac{2}{3}}\omega^{\frac{1}{2}}v^{-\frac{1}{6}}C}$$

where $i$, $i_k$, $n$, $F$, $A$, $D_O$, $\omega$, $v$, and $C$ indicate the measured current, kinetic current, electron transfer number, Faraday constant (96,485 C $mol^{-1}$), geometric disk area (0.196 $cm^2$), diffusion coefficient of $O_2$ ($1.86\times10^{-5}$ $cm^2$ $s^{-1}$), rotation speed (rad $s^{-1}$), $O_2$ kinematic viscosity ($1.008\times10^{-2}$ $cm^2$ $s^{-1}$), and $O_2$ concentration ($1.21\times10^{-6}$ mol $cm^{-3}$), respectively.[36] All ORR RDE testing was also conducted on the polished GC tip RDE as a control. Geometric current densities and potentials are reported as the mean ± SD ($n$ = 3).

**Electrical Impedance Spectroscopy (EIS).** EIS measurements were made at 0 V versus Ag/AgCl with a 10-mV potential amplitude from 50 kHz to 0.1 Hz in $N_2$-saturated 0.1 M KOH to measure solution resistance ($R_s$) and check electrode impedance before further ORR testing. To measure the electrode double-layer capacitance ($C_{dl}$) and charge transfer resistance ($R_{ct}$), EIS was performed at -0.2 V versus Ag/AgCl in $O_2$-saturated 0.1 M KOH. The Nyquist plots for each EIS spectra were fit with an equivalent Randles circuit (**Figure 3.b inset**) using PSTrace software and tabulated results are given in **Table S9**.

**Electrochemical stability.** The stability of NT-3DFG during ORR was investigated through chronoamperometry. NT-3DFG independently synthesized at 1100 °C for 120 min were held at 0.5 V vs. RHE in $O_2$-saturated 0.1 M KOH and current was measured as a function of time for 15 hours.

**Calculation Details.** Density Functional Theory (DFT) calculations were performed using the GPAW package with the BEEF-vdW exchange-correlation functional using the Atomic Simulation Environment (ASE). A Monkhorst-Pack type $k$-point grid of size 1 × 1 × 4 was used for AC and ZZ edge-terminated nanoribbon and core electrons were described using the Projector Augmented Wave Function (PAW). All calculations were performed with a grid spacing of 0.18 Å and converged with a force criterion threshold of 0.05 eV Å$^{-1}$. The free energies of oxygen intermediates were calculated using DFT at a potential of 0 V versus the RHE and at standard conditions by incorporating entropy contributions and zero-point energy. We plot free energy landscapes for high activity sites in **Figure S17**. The effect of potential $U$ is incorporated by shifting the free energy of electrons by $-eU$. Calculations for all investigated edge configurations are shown in **Figure S18**. In this work, we additionally leverage error estimation capabilities within the BEEF-vdW XC-functional, which allows us to compute an ensemble of energies on the basis of Bayesian statistics. The error estimation capability has previously been used to describe the robustness in predicting magnetic ground states, reaction mechanism pathways, heterogeneous catalytic activity, and solid-state properties.

**Catalyst Active Site Characterization through Pourbaix Diagrams.** The surface Pourbaix diagram represents the thermodynamically stable state of the catalyst surface in electrochemical systems as a function of pH and electrode potential U. A generalized representation of the adsorption of intermediates (denoted as $O_mH_n^*$) relevant for hydrogen peroxide formation using NT-3DFG can be expressed as a concerted proton-coupled electron transfer (PCET) reaction:

$$S - O_mH_n^* + (2m - n)(e^- + H^+) \rightarrow S^* + mH_2O \quad \text{(eqn. 1)}$$

where, an edge site is denoted as $S$, and * represents an adsorption site in its pristine state. $m$ and $n$ are the number of oxygen and hydrogen atoms in the adsorbate respectively. It is worth pointing out that in this study, the considered edge terminations groups are H*, O* and OH*. The associated free energy change can be computed as:

$$\Delta G(U, pH) = G_{S^*} + mG_{H_2O} - G_{S-O_mH_n^*} - (2m - n)(\frac{1}{2}G_{H_2} - eU_{SHE} - 2.303\, k_BT\, pH) \quad \text{(eqn. 2)}$$

where $e$ is the electron charge. $k_B$ is the Boltzmann constant, and $U_{SHE}$ is the potential relative to the standard hydrogen electrode (SHE). This allows us to derive relations between the electrode potential and $pH$ when $\Delta G(U, pH) = 0$ (equilibrium) for a range of adsorbates on an edge sites, which form the phase boundaries. The surface Pourbaix diagram (**Figure S19, and Figure S20**) is constructed by choosing the appropriate lowest energy state at each set of conditions of $U$ and $pH$. Free energies are determined from calculated adsorption energies, $\Delta E$, and by correcting for entropy changes, $\Delta S$, and zero-point energy change, $\Delta ZPE$, which we assume do not change significantly with temperature, by $\Delta G = \Delta E - T\Delta S + \Delta ZPE$.

**Prediction Confidence of Surface Pourbaix Diagrams.** Pourbaix diagrams typically exhibit sharp phase boundaries and depict only the minimum Gibbs free energy (most thermodynamically stable) surface states over a range of operating potentials ($U$) and $pH$ values. We leverage error-estimation capabilities within the BEEF-vdW exchange correlation functional, which allows us to apply statistical tools on an ensemble of derived Pourbaix diagrams (**Figure S21, and Figure S22**) to obtain a measure of the confidence (c-value)[50] in a predicted stable surface state, a metric that quantifies the degree of agreement between functionals. In this context, the c-value is defined as the fraction of the ensemble that is in agreement with the hypothesis of the optimal BEEF-vdW (best-fit) functional. This captures the level of agreement between functionals at the GGA-level as to the most energetically favorable surface state as a function of potential and $pH$.

**Energy Loss Near Edge Structure (ELNES) Simulations.** The theoretical ELNES spectra were simulated using ab initio real space full multiple scattering Green's function techniques as implemented in FEFF9 program.[64, 65] The self-consistent field (SCF) calculations were performed considering the Hedin-Lundqvist self-energy model as exchange and correlation potential which described the atomic background as well as the energy loss near edge fine structure. To estimate the threshold energy of the carbon *K*-edge, the core state was treated with RPA-screened core-hole during the SCF loop. The full multiple scattering Green's function was calculated with a cluster radius large enough to specify about 50 atoms within the sphere. Calculations assumed a 300 keV beam hitting the sample along the normal to the surface. The detector was set with a collection semiangle, $\beta = 5.2$ mrad and convergence semiangle, $\alpha = 1.87$ mrad. The cross section was then integrated over the detector aperture using a 10 x 10 x 1 grid describing impulse transfer q values allowed by $\alpha$ and $\beta$. An instrumental broadening of 0.5 eV was introduced as the imaginary energy shift of the Green's function which also corrected the core-hole life-time. For various structure models, the DFT optimized structures were used in the calculations.

## Associated Content

### Supporting Information

The Supporting Information is available free of charge on the XX website at DOI:

Figure S1-S22 (NT-3DFG growth kinetics, DLA simulation, effects of NT-3DFG synthesis time, additional Raman analysis, tunable morphology of NT-3DFG edges, double-layer capacitance

characterization, ORR cyclic voltammetry response, Tafel slope analysis, electrical characterization, Koutecky-Levich analysis, chronoamperometry during ORR, morphology before & after ORR, Raman spectroscopy before & after ORR, simulated EELS spectra for ZZ edge atoms, distributions of coefficients in structure-property relationship, NT-3DFG edge detection, free energy landscape for high activity sites, full theoretical activity volcano, surface state of AC edge sites, surface state of ZZ edge sites, c-values associated with AC surface state predictions, c-values associated with ZZ surface state predictions); Table S1-14 (data summaries for Raman analysis, dual-laser Raman analysis, double-layer capacitance characterization, electrical characterization, EIS, $H_2O_2$ selectivity, Raman analysis after ORR, XPS analysis before & after ORR, C1s & O1s chemical states of NT-3DFG after ORR, explored ZZ & AC active sites with local coordination environment) (PDF)

## Author Information

### Corresponding Authors

*E-mail: tzahi@andrew.cmu.edu

*E-mail: venkatv@andrew.cmu.edu

### ORCID

Daniel J. San Roman: 0000-0003-3133-1446

Dilip Krishnamurthy: 0000-0001-8231-5492

Raghav Garg: 0000-0002-3501-6892

Hasnain Hafiz: 0000-0002-0202-794X

Venkatasubramanian Viswanathan: 0000-0003-1060-5495

Tzahi Cohen-Karni: 0000-0001-5742-1007

### Author Contributions

D.S.R. and D.K. contributed equally to this work.

### Notes

The authors declare no competing financial interests.

## Acknowledgements

T.C-K. gratefully acknowledges funding support from the National Science Foundation under Award No. CBET1552833 and the Office of Naval Research under Award No. N000141712368. D.K. and V.V. gratefully acknowledge funding support from the National Science Foundation under Award No. CBET1554273. H. H. acknowledges support from the U.S. Department of Energy's Office of Energy Efficiency and Renewable Energy (EERE) under the Fuel Cell Technologies Office (FCTO) under Award Number DE-EE0008076. We thank S. Lister and X. Xu for assistance with electrochemistry experiments and thank D. Flaherty for assistance with experimental EELS measurements. We also acknowledge support from the Carnegie Mellon University Nanofabrication Facility and Department of Materials Science and Engineering Materials Characterization Facility (MCF-677785).

## Supporting Information for

# Engineering Three-Dimensional (3D) Out-of-Plane Graphene Edge Sites for Highly Selective Two-Electron Oxygen Reduction Electrocatalysis

*Daniel San Roman*[1^], *Dilip Krishnamurthy*[2^], *Raghav Garg*[1], *Hasnain Hafiz*[2], *Michael Lamparski*,[4] *Noel T. Nuhfer*[1], *Vincent Meunier*,[4] *Venkatasubramanian Viswanathan*[2*] and *Tzahi Cohen-Karni*[1,3*]

[1] *Department of Material Science and Engineering, Carnegie Mellon University, Pittsburgh, Pennsylvania, 15213, USA.*

[2] *Department of Mechanical Engineering, Carnegie Mellon University, Pittsburgh, Pennsylvania, 15213, USA.*

[3] *Department of Biomedical Engineering, Carnegie Mellon University, Pittsburgh, Pennsylvania, 15213, USA.*

[4] *Department of Physics, Applied Physics, and Astronomy, Rensselaer Polytechnic Institute, Troy, New York 12180, USA*

^ *Equal Contribution*

* *Corresponding author*

## Table of Contents

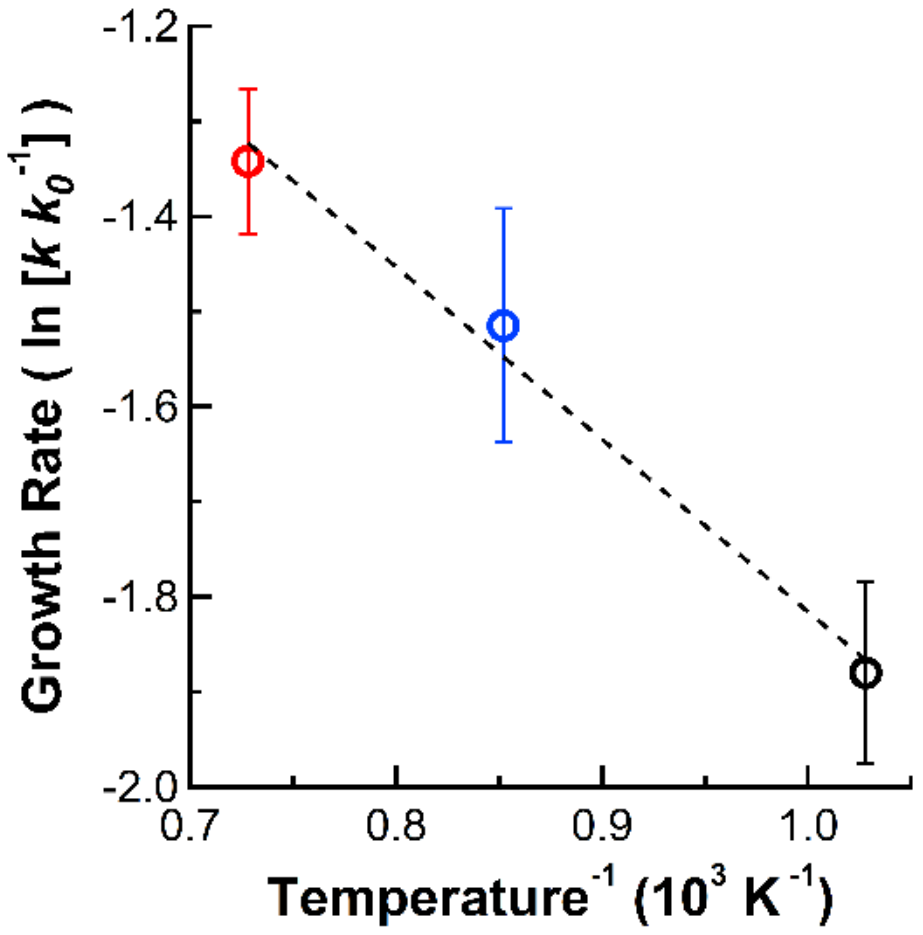

**Figure S1. NT-3DFG growth kinetics.** Arrhenius plot for NT-3DFG growth at 700, 900, and 1100 °C (**black**, **blue**, and **red**, respectively.) Results are presented as mean ± standard deviation (SD) with sample size (*n)* of 3.

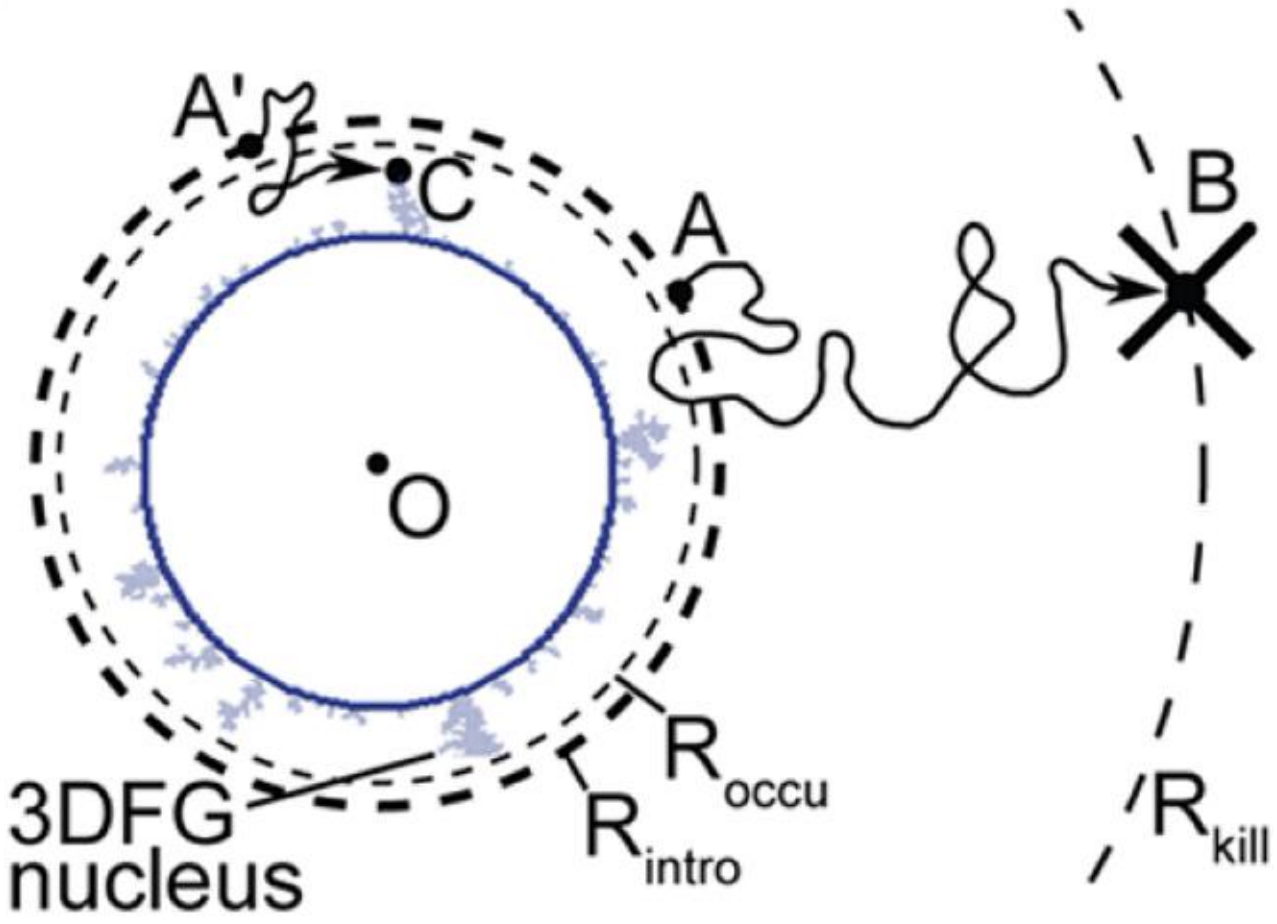

**Figure S2. Diffusion-limited aggregation simulation for NT-3DFG.** Illustration of a single nucleation event in a simulation of diffusion-limited aggregation as shown in previous literature.[ref] O, $R_{intro}$, and $R_{occu}$ represent the origin, surface radius of dimer introduction, and maximal surface radius of currently occupied site, respectively. A and B (or $R_{kill}$), and A’ and C correspond to unsuccessful and successful nucleation events, respectively.

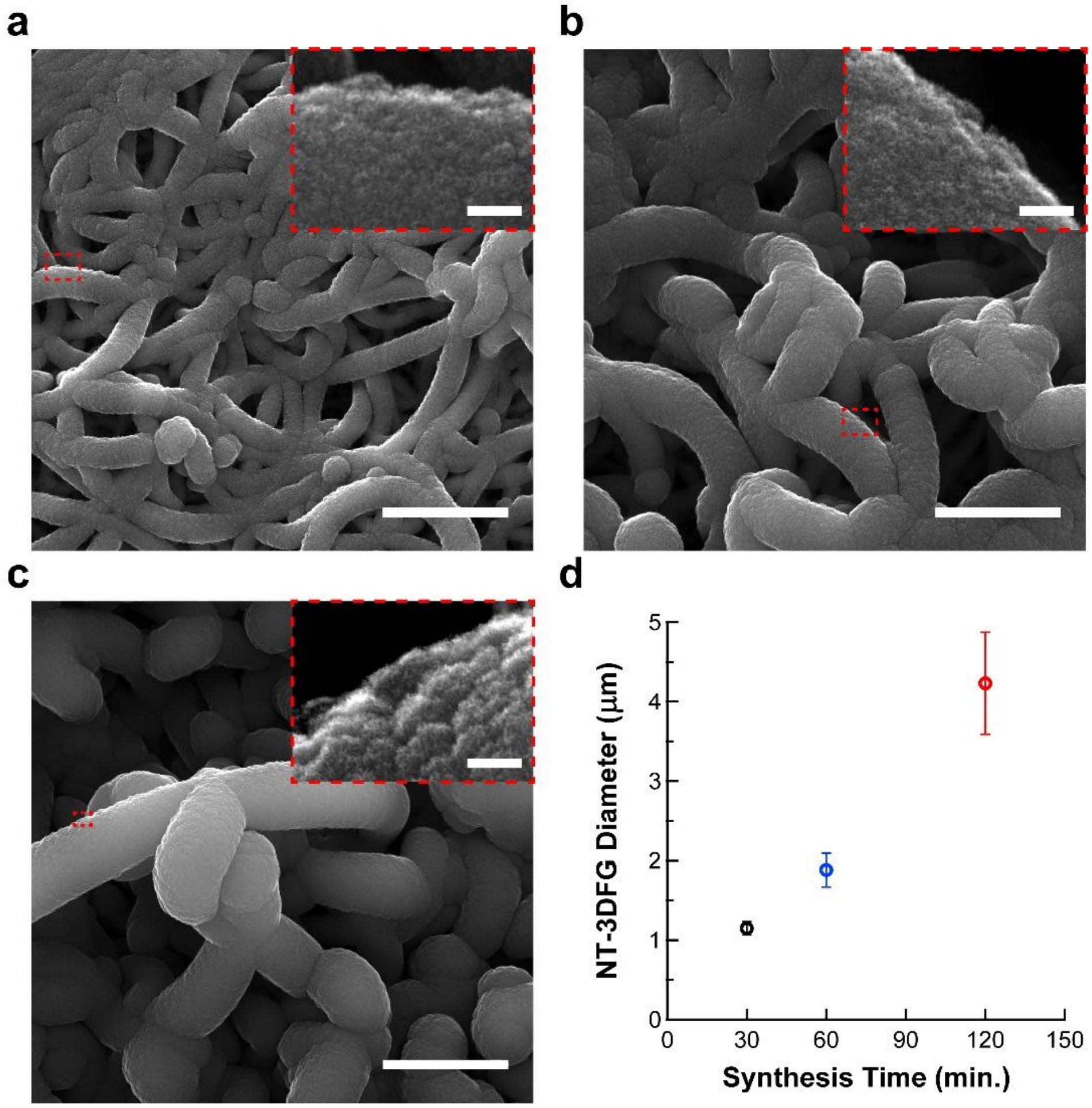

**Figure S3. Effects of synthesis time. (a-c)** SEM images of NT-3DFG synthesized for 30, 60, 120 minutes at 1100 °C, respectively. Scale bars: 5, 5, and 10 µm, respectively. Insets represented by red dashed boxes show out-of-plane graphene on SiNW. Inset scale bars: 300 nm. **(d)** NT-3DFG diameter as a function of synthesis time at 1100 °C. Results are presented as mean ± SD ($n$ = 3).

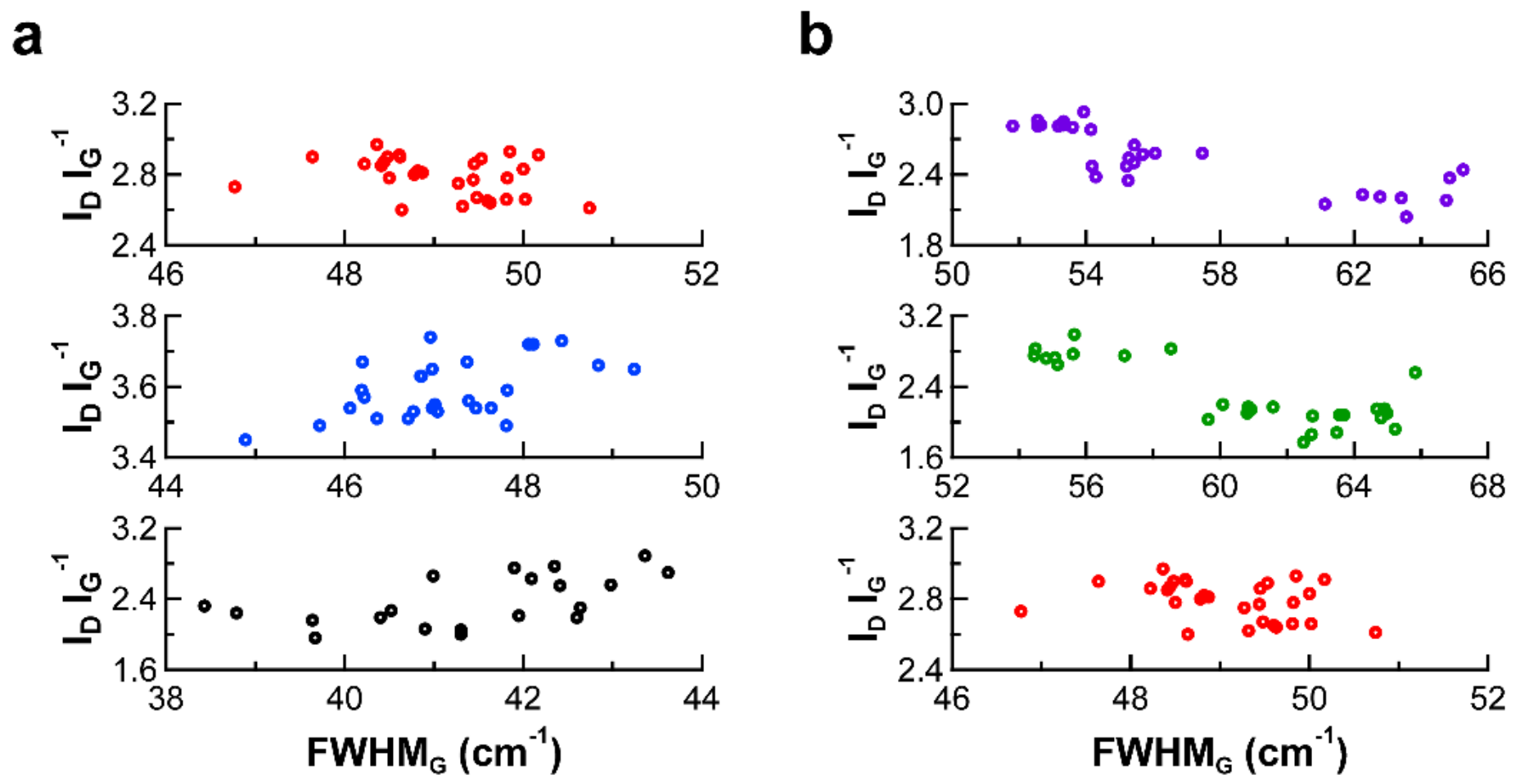

**Figure S4. Additional Raman analysis.** $I_D\ I_G^{-1}$ as a function of FWHM$_G$ for NT-3DFG synthesized at **(a)** 700 (**black**), 900 (**blue**), and 1100 °C (**red**) for 30 min; and **(b)** 1100 °C for 30 (**red**), 60 (**green**), and 120 min (**purple**). $n$ = 3 for both **a** and **b**.

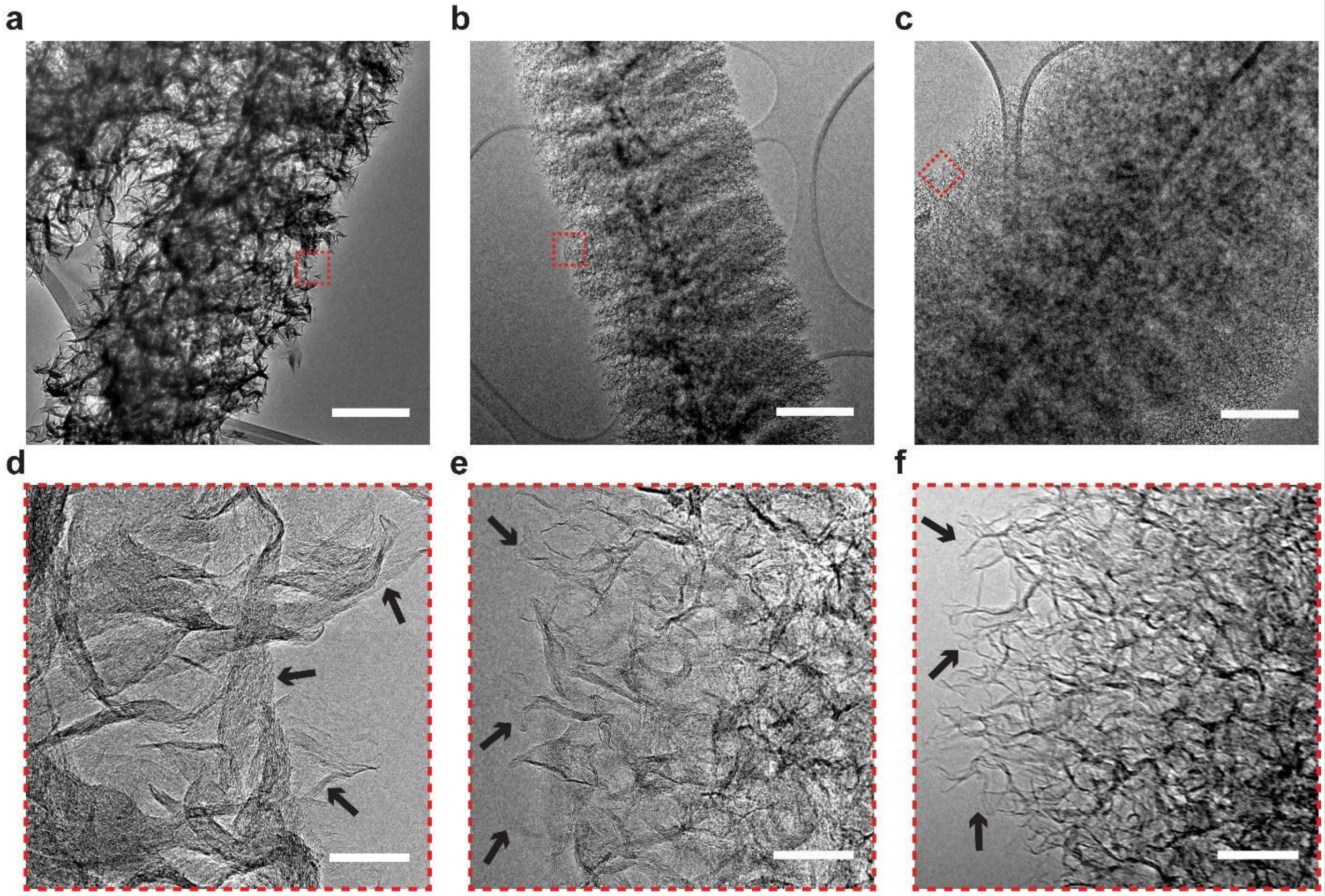

**Figure S5. Tunable morphology of NT-3DFG edges. (a-c)** Representative HRTEM images of NT-3DFG synthesized for 30 minutes at 700, 900, and 1100 °C, respectively. Scale bars: 250 nm. **(d-e)** Expanded view from **a-c**, respectively, represented by red boxes, show edge density. Black arrows point to terminations of 3DFG flakes that approach single layer graphene in nature at higher synthesis temperatures. Scale bars: 20 nm

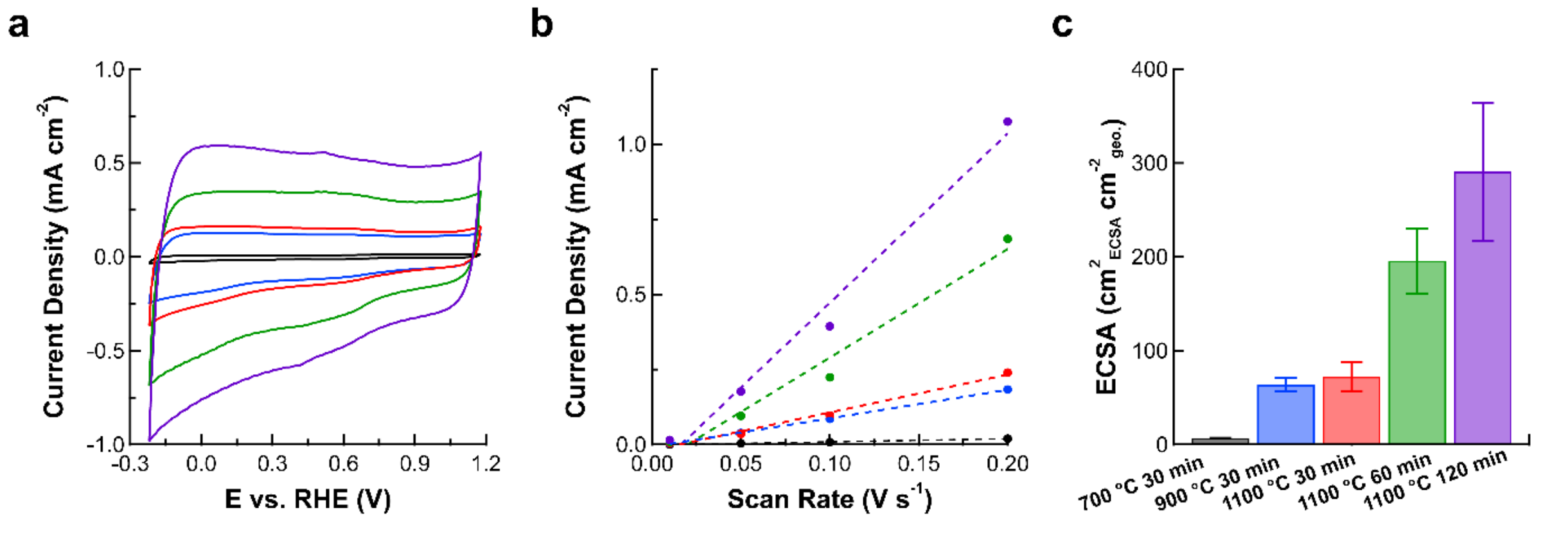

**Figure S6. Double-layer capacitance characterization. (a)** Representative cyclic voltammograms (CV) ($n$ = 3) acquired in $N_2$-saturated 0.1 M KOH at scan rate of 100 mV $s^{-1}$ for NT-3DFG synthesized at 700 °C for 30 min (**black**), 900 °C for 30 min (**blue**), 1100 °C for 30 min (**red**), 1100 °C for 60 min (**green**), and 1100 °C for 120 min (**purple**). **(b)** Current density measured at 0.95 V vs. RHE as a function of scan rate. **(c)** Electrochemical surface area values (ECSA) for NT-3DFG as measured by the differential double-layer capacitance. Results are presented as mean ± SD ($n$ = 3).

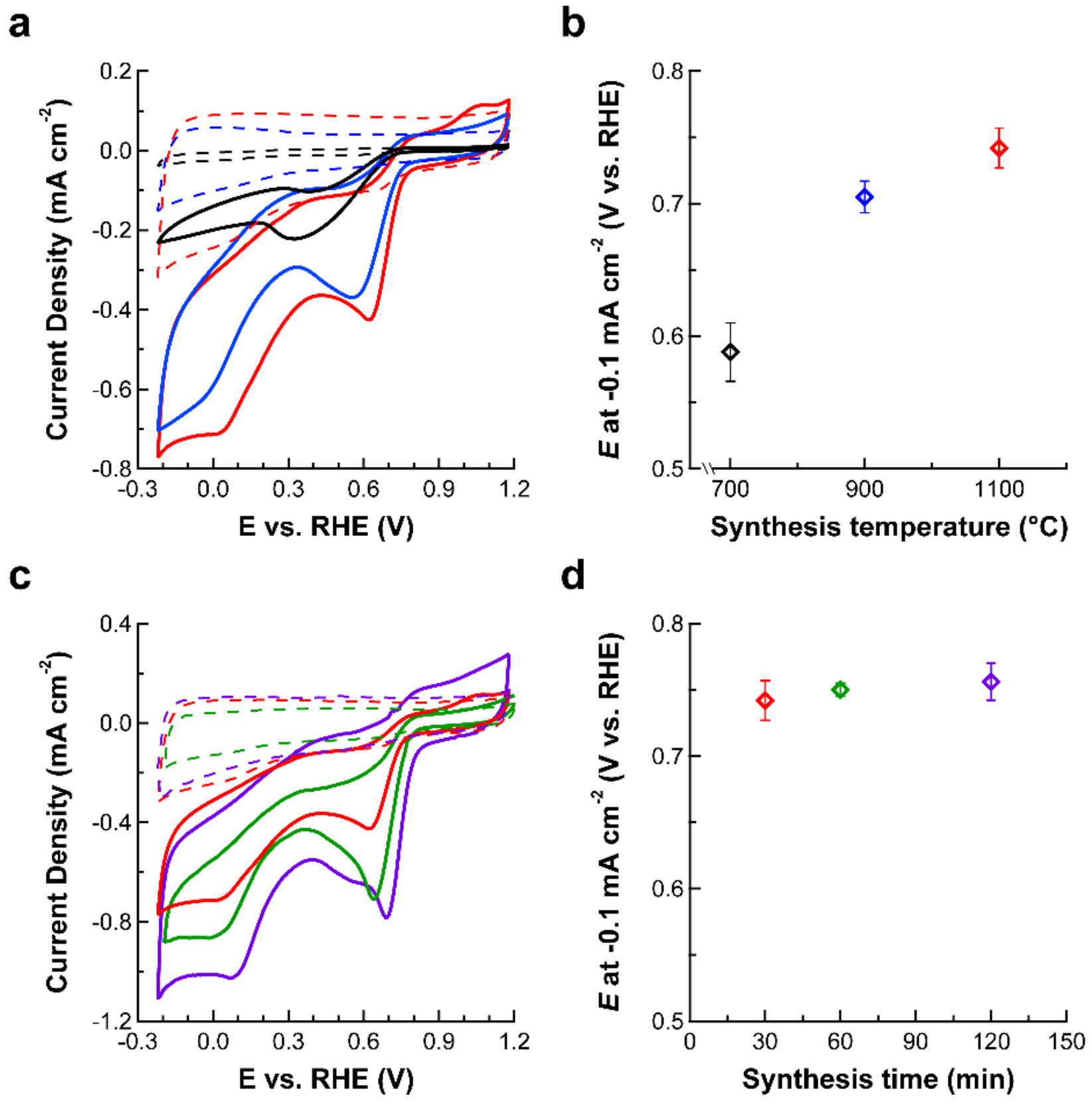

**Figure S7. Oxygen reduction reaction (ORR) cyclic voltammetry response. (a)** Representative cyclic voltammograms (CV) (*n* = 3) for NT-3DFG synthesized for 30 minutes at 700 °C (**black**), 900 °C (**blue**), and 1100 °C (**red**). **(b)** Potential at -0.1 mA $cm^{-2}$ after baseline-current subtraction for NT-3DFG synthesized for 30 minutes at 700 °C (**black**), 900 °C (**blue**), and 1100 °C (**red**). **(c)** Representative CV (*n* = 3) of NT-3DFG synthesized at 1100 °C for 30 min (**red**), 60 min (**green**), and 120 min (**purple**). **(a, c)** CV was conducted in $N_2$-saturated (**dash**) and $O_2$-saturated (**solid**) 0.1 M KOH at a scan rate of 50 mV $s^{-1}$. **(d)** Potential at -0.1 mA $cm^{-2}$ after baseline-current subtraction for NT-3DFG synthesized at 1100 °C for 30 min (**red**), 60 min (**green**), and 120 min (**purple**). **(b, d)** Results are presented as mean ± SD (*n* = 3).

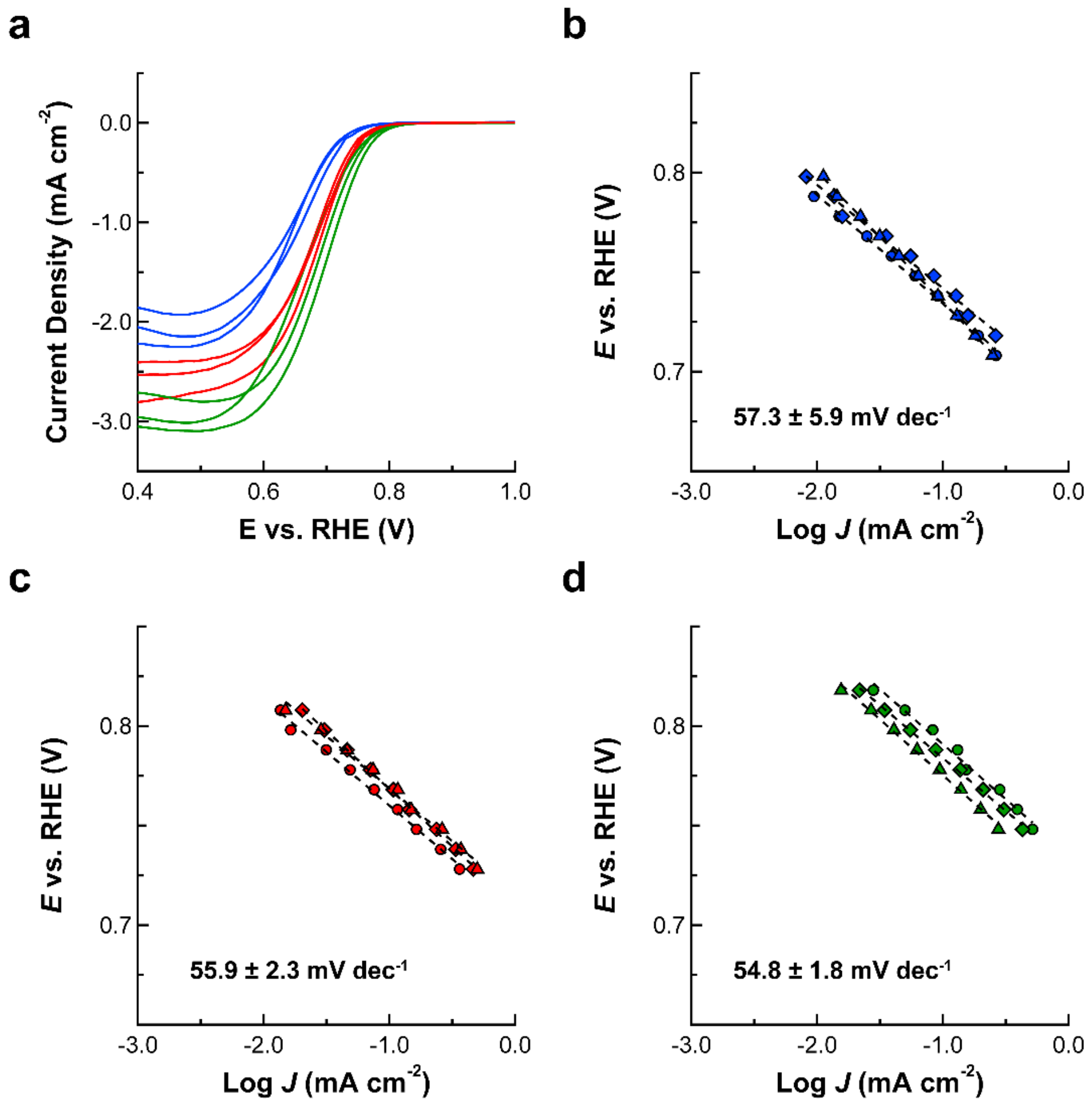

**Figure S8. Tafel slope analysis. (a)** Rotating disk electrode linear sweep voltammograms (LSV) for three independent NT-3DFG samples synthesized at 700 °C for 30 minutes (**blue**), 1100 °C for 30 min (**red**), and 1100 °C for 120 min (**green**) in $O_2$-saturated 0.1 M KOH at a scan rate of 10 mV $s^{-1}$ and rotational speed of 1600 rpm. **(b-d)** Tafel plots denoted by **circles**, **diamonds**, and **triangles** for LSV curves shown in **a** for NT-3DFG synthesized at 700 °C for 30 minutes (**b**), 1100 °C for 30 min (**c**), and 1100 °C for 120 min (**d**). The fit for each Tafel plot is shown as a dashed line. The Tafel slopes for each plot are presented as mean ± SD (*n* = 3).

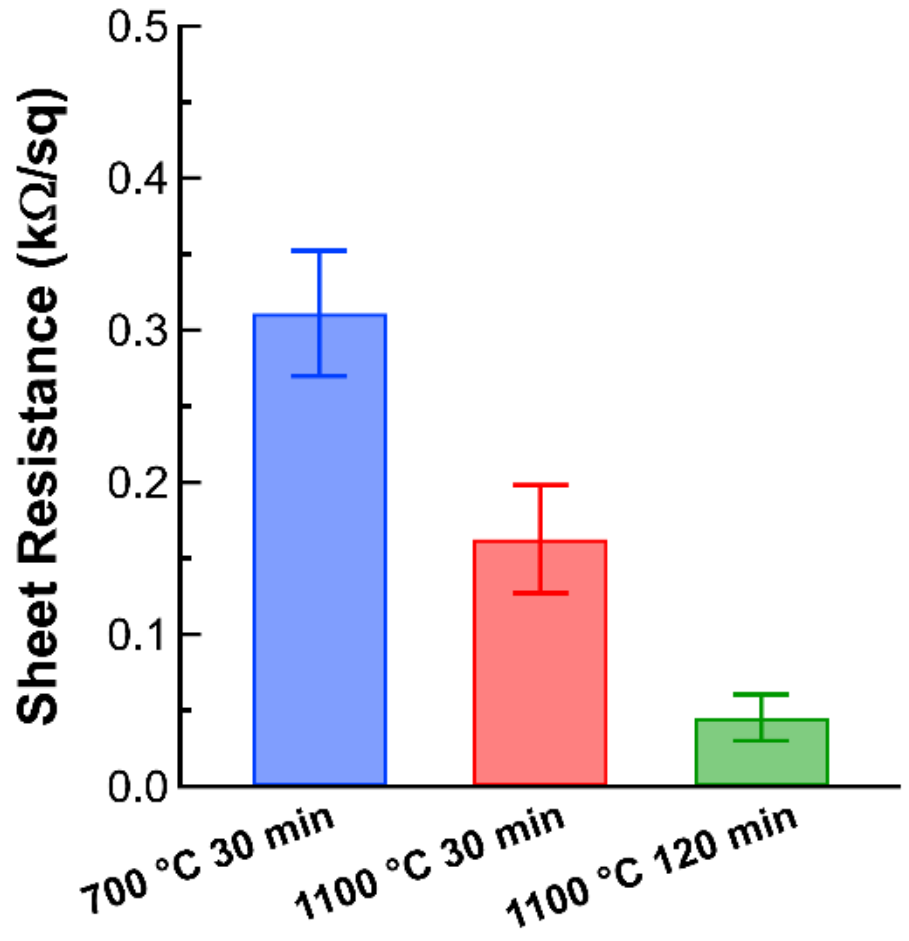

**Figure S9. Electrical characterization of NT-3DFG catalysts.** Sheet resistance of NT-3DFG synthesized at 700 °C for 30 minutes (**blue**), 1100 °C for 30 min (**red**), and 1100 °C for 120 min (**green**). Results are presented as mean ± SD ($n$ = 3).

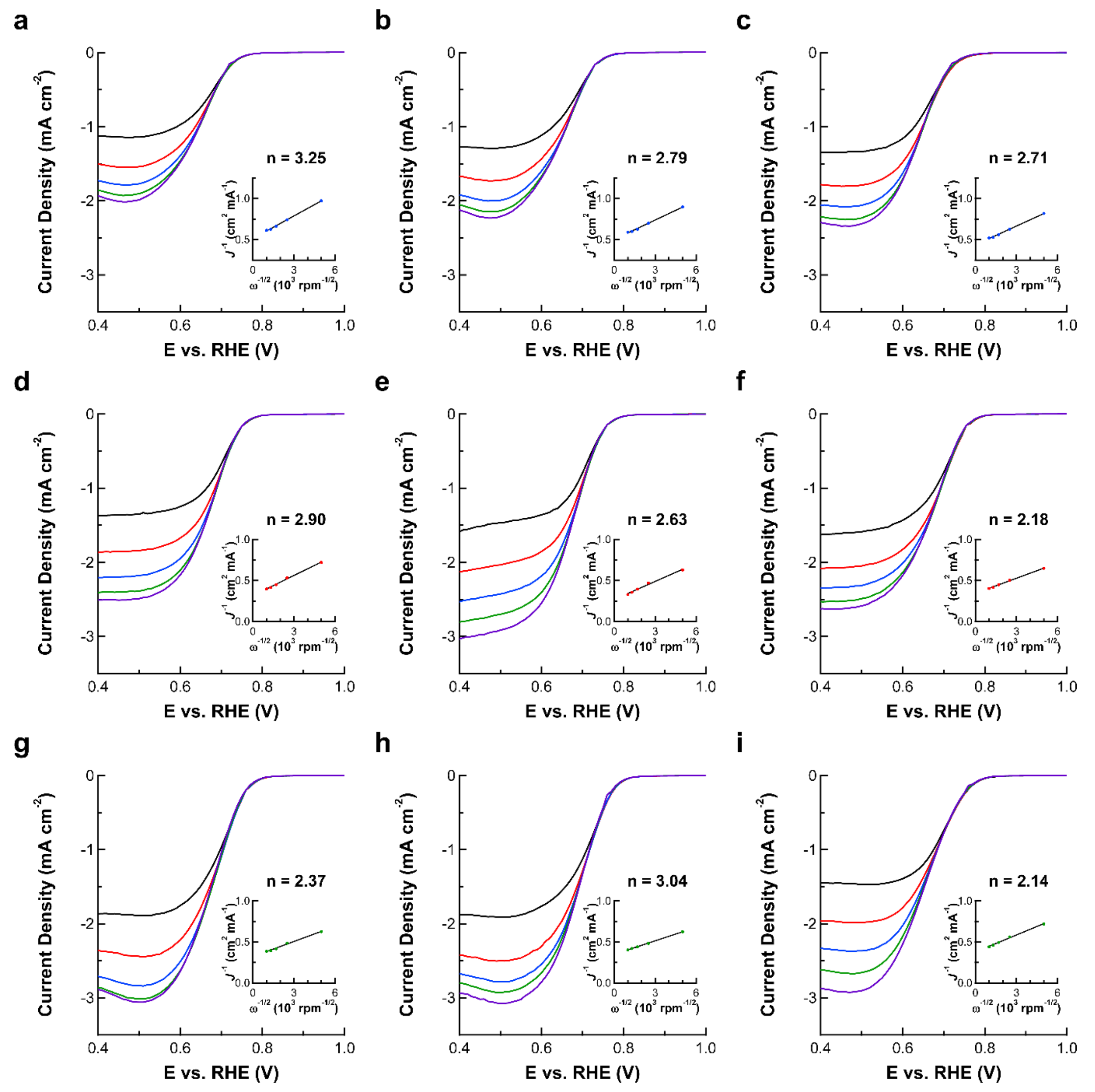

**Figure S10. Koutecky-Levich analysis. (a)** Rotating disk electrode linear sweep voltammograms (LSV) for three independent NT-3DFG samples synthesized at 700 °C for 30 minutes (**a-c**), 1100 °C for 30 min (**d-f**), and 1100 °C for 120 min (**g-i**) in $O_2$-saturated 0.1 M KOH at a scan rate of 10 mV $s^{-1}$ and rotational speeds of 400 rpm (**black**), 800 rpm (**red**), 1200 rpm (**blue**), 1600 rpm (**green**), and 2000 rpm (**purple**). Insets show inverse current densities at 0.5 V vs. RHE as a function of rotational speed with fitted slopes and recorded ORR electron transfer numbers (*n*).

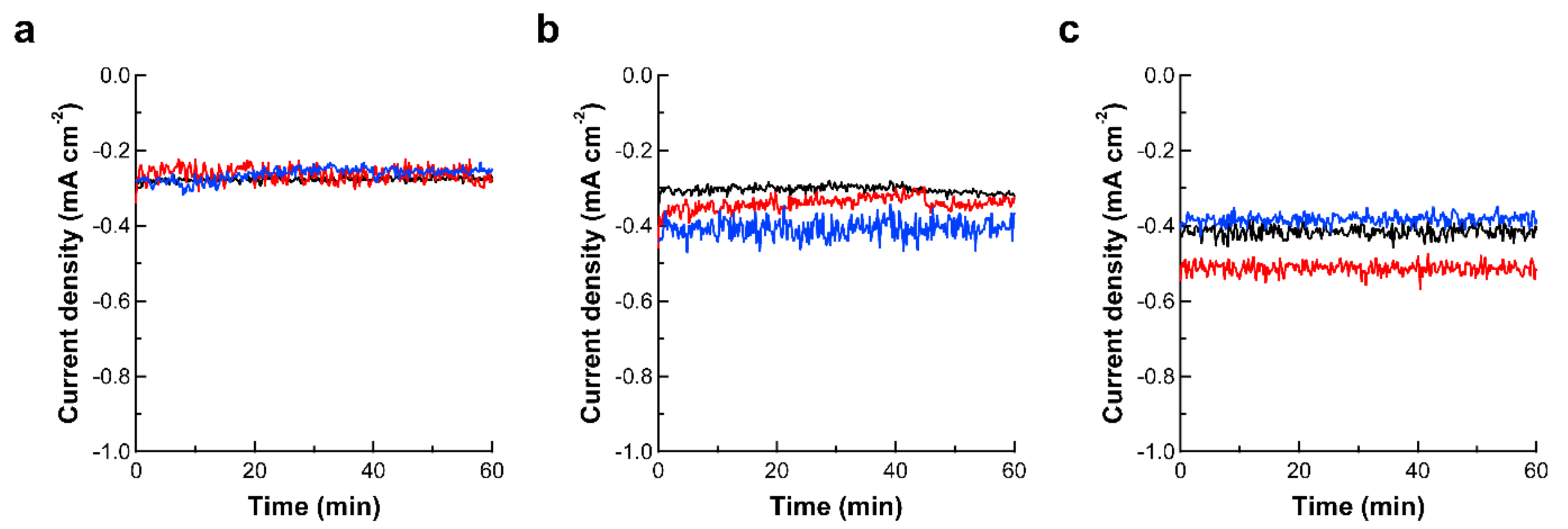

**Figure S11. Chronoamperometry (CA) during ORR.** CA curves for three (**red**, **black**, & **blue**) independent NT-3DFG electrodes synthesized at 700 °C for 30 minutes (**a**), 1100 °C for 30 min (**b**), and 1100 °C for 120 min (**c**). Testing was conducted for 1 hour at 0.5 V vs. RHE in $O_2$-saturated 0.1 M KOH.

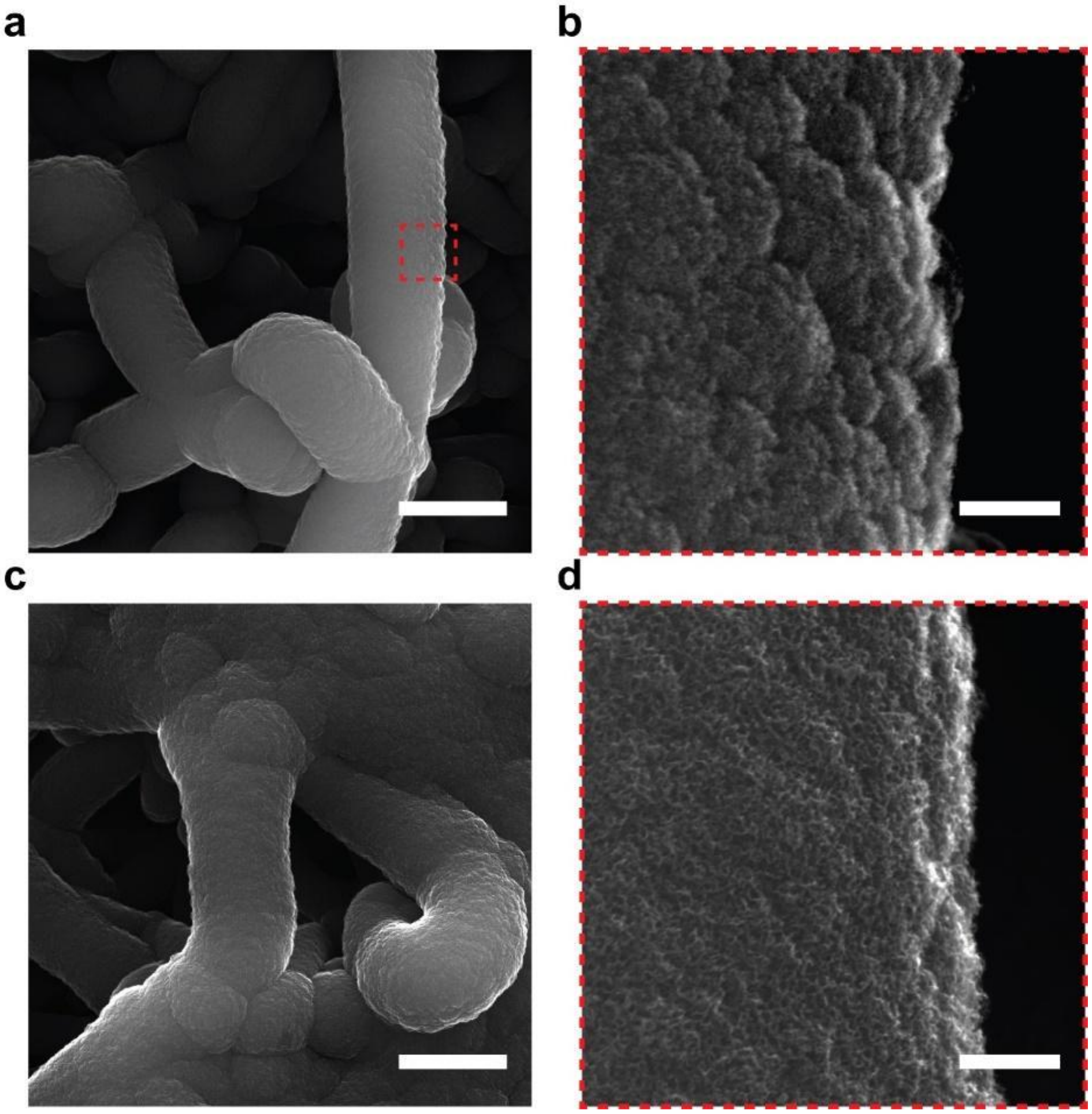

**Figure S12. Morphology of NT-3DFG before & after ORR. (a)** NT-3DFG before ORR testing. **(b)** Expanded view from **(a)** represented by red dashed boxes show edges of 3DFG. **(c)** NT-3DFG after ORR CV and $H_2O_2$ selectivity testing. **(d)** Expanded view from **(c)** represented by red dashed boxes show edges of 3DFG. Scale bars: 3 µm **(a,c)** and 300 nm **(b,d)**. NT-3DFG was synthesized at 1100 °C for 120 minutes.

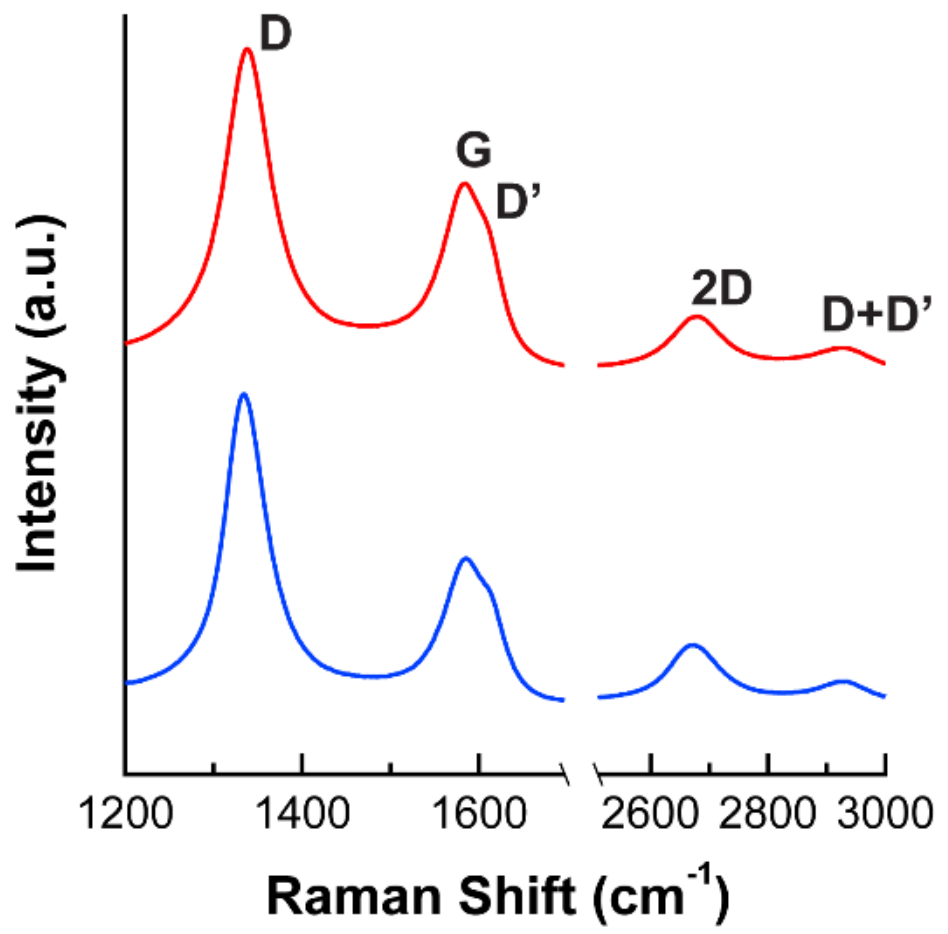

**Figure S13. Raman spectroscopy of NT-3DFG before & after ORR. (a)** Raman spectra of NT-3DFG before ORR testing (**blue**) and NT-3DFG after ORR CV and $H_2O_2$ selectivity testing (**red**.) NT-3DFG was synthesized at 1100 °C for 120 minutes.

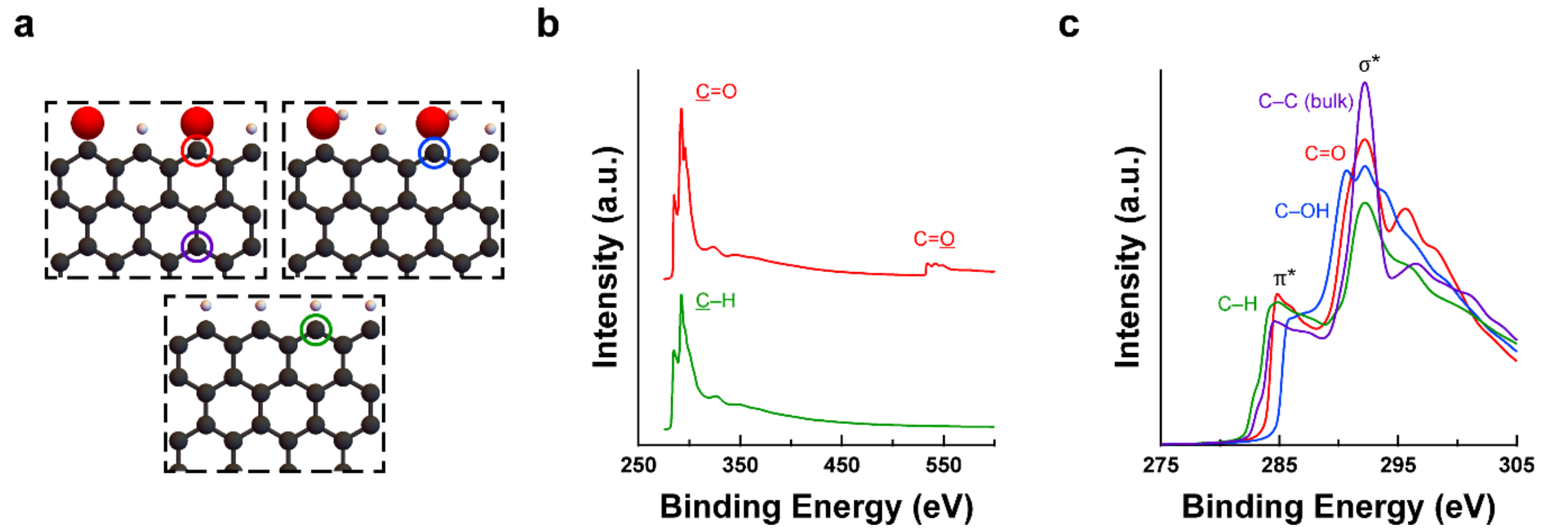

**Figure S14. Simulated EELS spectra for Zigzag (ZZ) edge atoms of NT-3DFG. (a)** ZZ edge terminations considered for ELNES calculations. O, H and C atoms are represented using red, silver and black spheres. **(b)** Theoretical EELS spectra associated with the C–H group (**green**) and C=O group (**red**). **(c)** Theoretical ELNES spectra for ZZ edge configurations with spectra colors corresponding to respective highlighted C atoms.

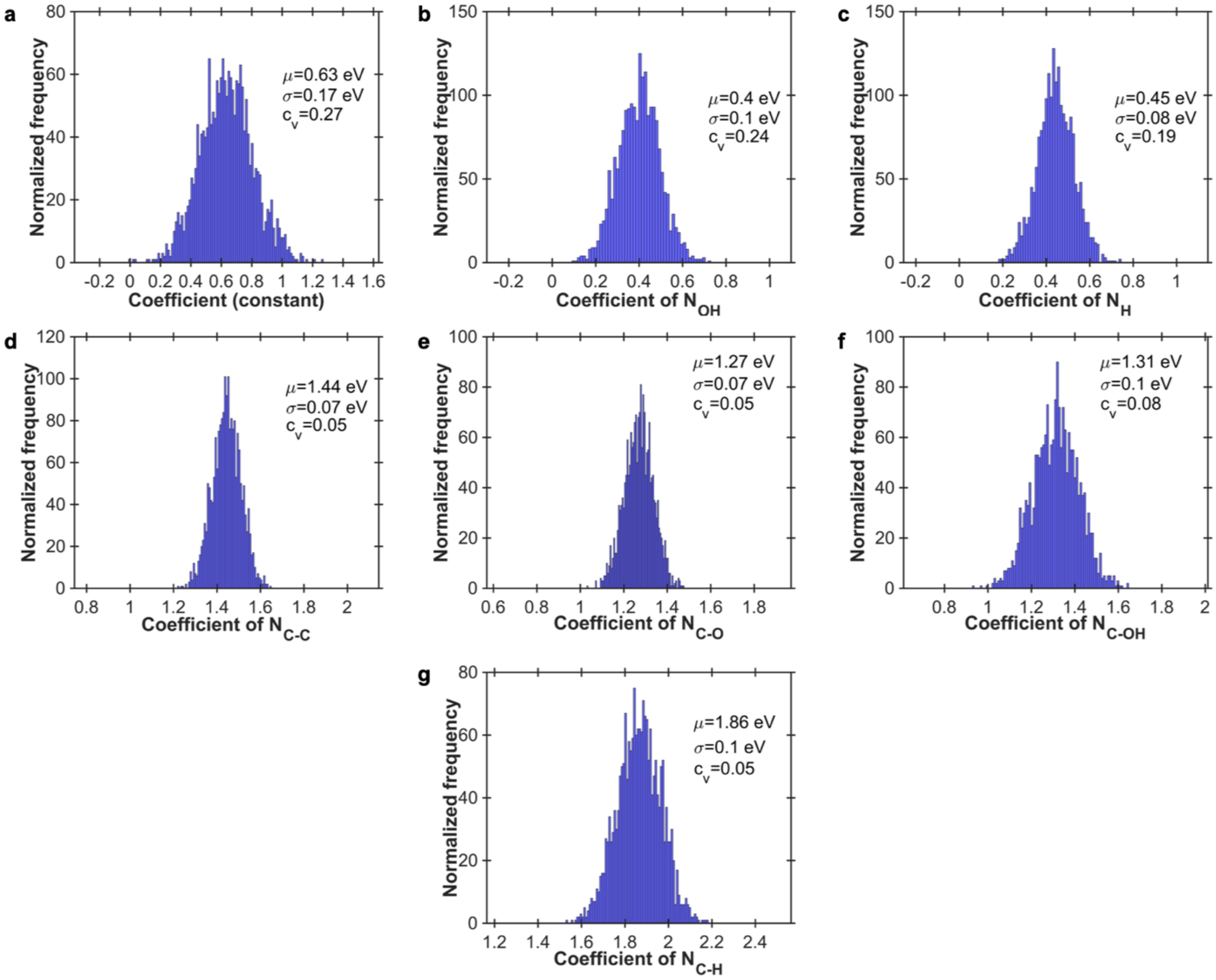

**Figure S15. Distributions of coefficients in the structure-property relationship**. Distributions obtained by constructing an ensemble of models using error estimation capabilities available within the BEEF-vdW exchange correlation functional. $\sigma$ represents the standard deviation, $\mu$ represents the mean, and $c_v$ represents the coefficient of variation with number of data points, $N_{ens}$, equal to 2000. We observe tight distributions for the coefficients based on the coefficient of variation, which indicates robustness of the constructed structure-activity relationship.

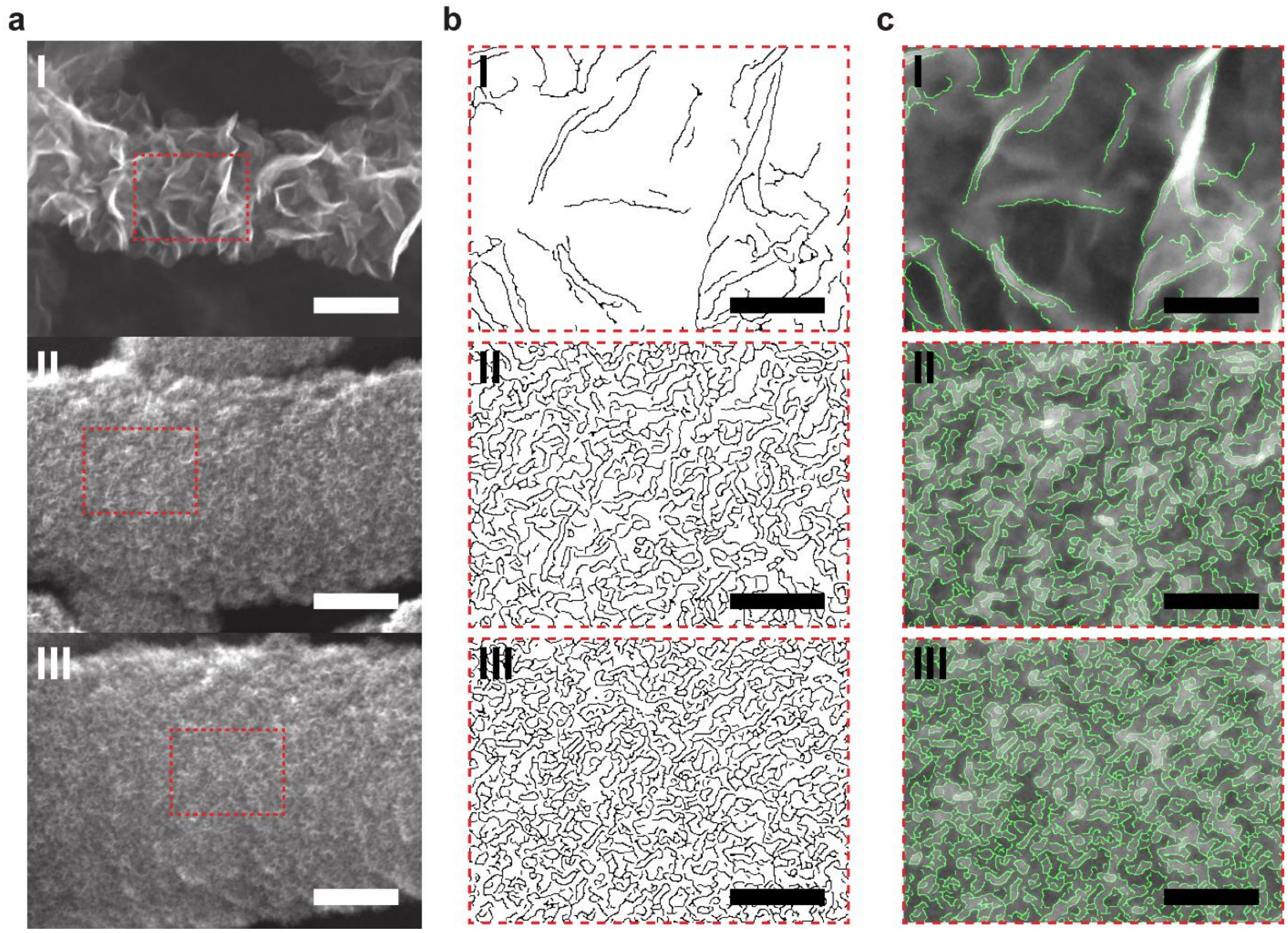

**Figure S16. Edge Detection of NT-3DFG. (a)** Representative SEM images of NT-3DFG synthesized for 30 minutes at 700, 900, and 1100 °C, (**a.I-III**, respectively.) Scale bars: 300 nm. **(b)** Expanded views of images (**a.I-III**) denoted by dashed red lines showing respective edge maps (**b.I-III**) of NT-3DFG. Scale bars: 100 nm. **(c)** Superimposed edge maps (**b.I-III**, green) on corresponding SEM images (**a.I-III**, greyscale). Scale bars: 100 nm.

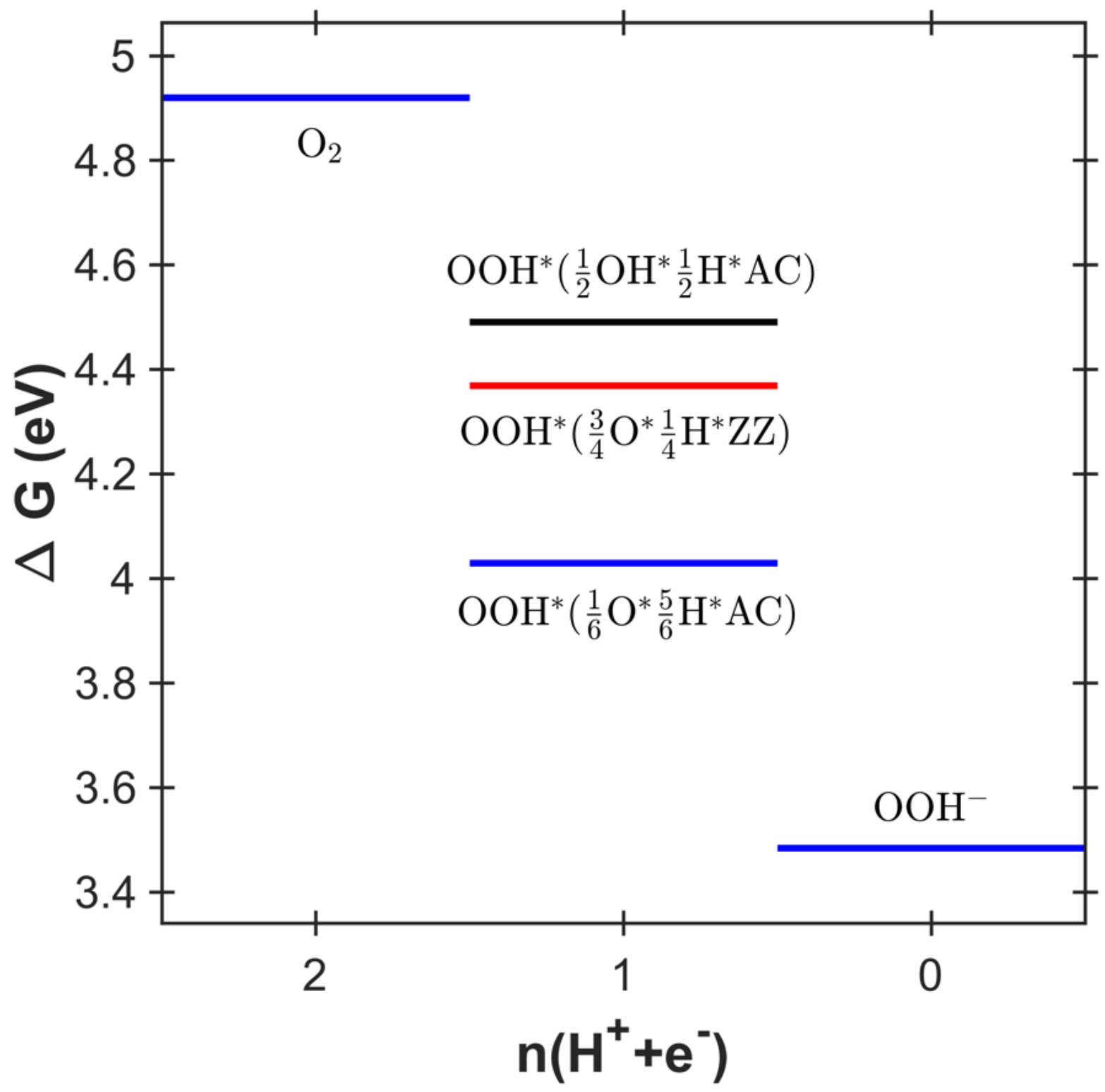

**Figure S17. Free energy landscape for the active sites** computed to exhibit the highest activity: $\frac{3}{4}O^*\frac{1}{4}H^*$ ZZ, $\frac{1}{6}O^*\frac{5}{6}H^*$ AC, and $\frac{1}{2}OH^*\frac{1}{2}H^*$ AC. The reference ($\Delta G = 0$) for the plotted free energies is the level of water. The x-axis refers the concerted addition of the number ($n$) of protons and electrons for the formation of hydrogen peroxide associated with each plotted free energy. We note that the dissociated form ($OOH^-$) has a lower free energy than that of $H_2O_2$ (pKa=11.7) under alkaline conditions (pH=13).

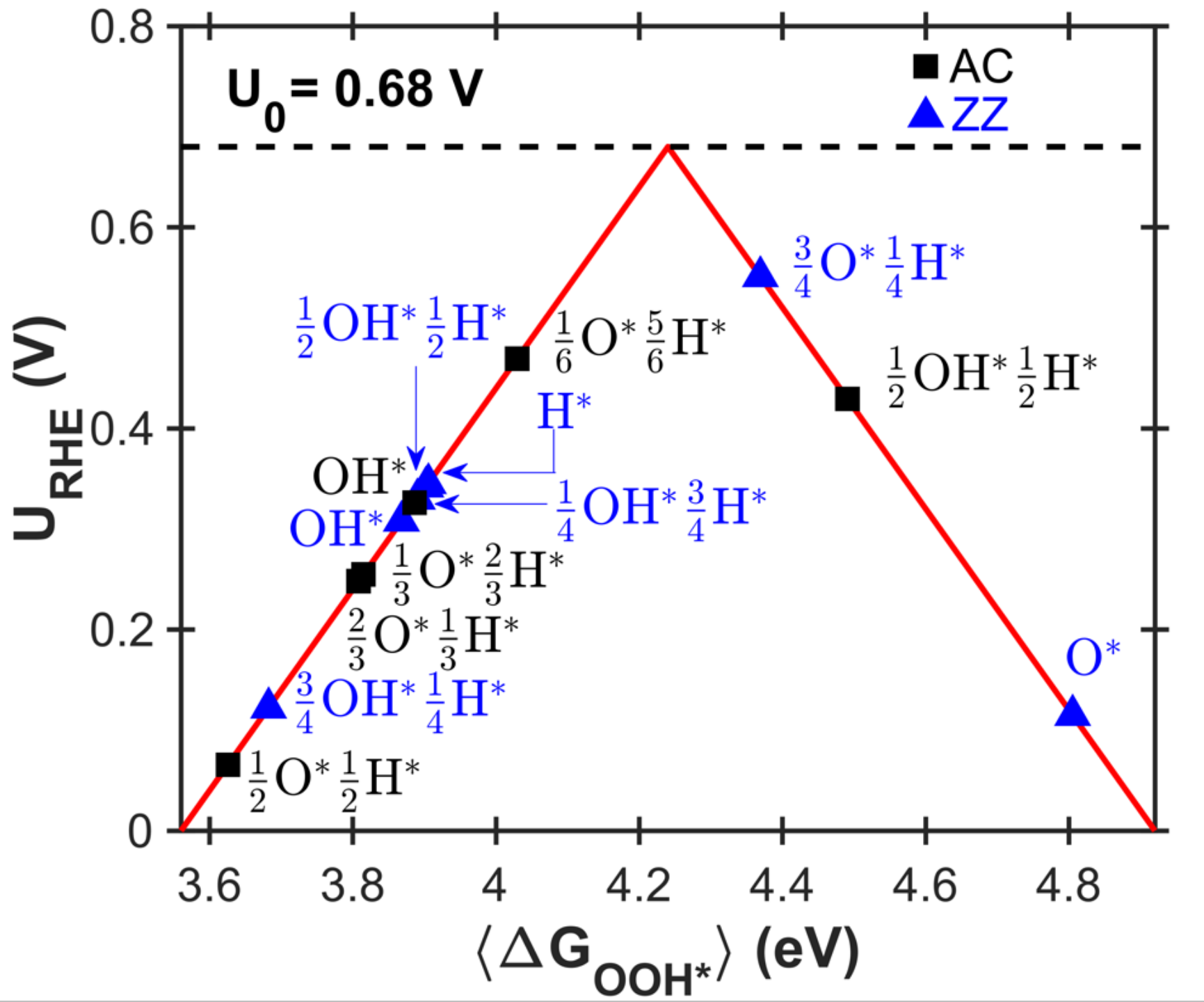

**Figure S18. Theoretical activity volcano for 2-electron ORR at graphene edge sites**. Theoretical activity volcano for 2-electron oxygen reduction showing activity predictions of the most stable active sites on various edge site termination configurations.

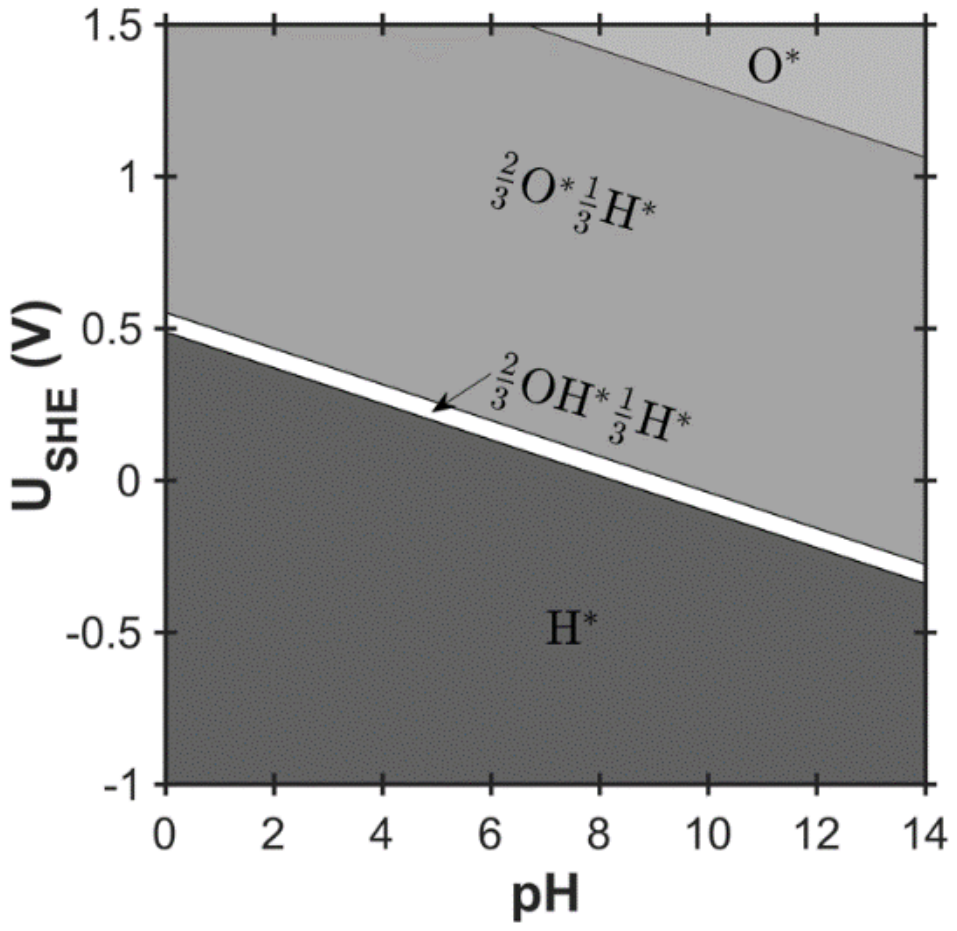

**Figure S19. Determining the state of the surface of AC edge sites.** Pourbaix diagram showing the most energetically favorable surface species configuration of AC edge sites, as a function of reaction conditions of electrode potential and pH, predicted from the optimal BEEF-vdW exchange correlation functional. Our analysis suggests that for sites that exhibit activity below ≈0.5 V, the surface is hydrogen saturated.

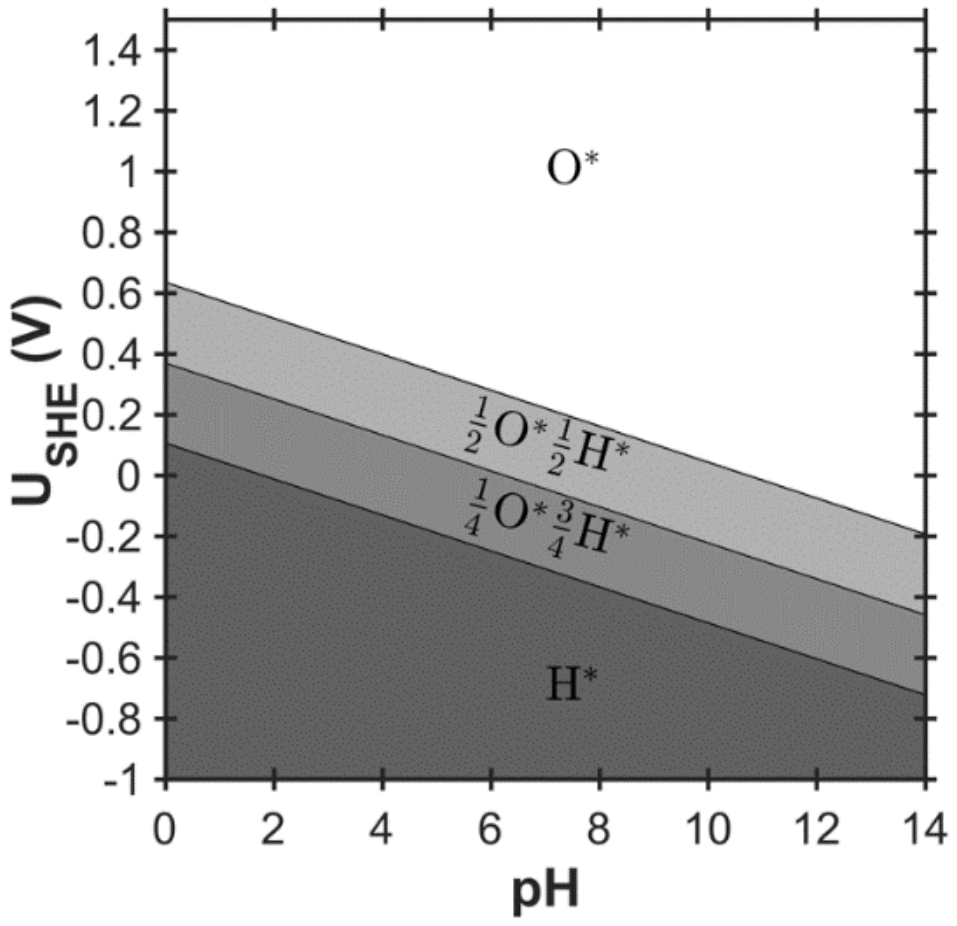

**Figure S20. Determining the state of the surface of ZZ edge sites.** Pourbaix diagram showing the most energetically favorable surface species configuration of ZZ edge sites, as a function of reaction conditions of electrode potential and pH, predicted from the optimal BEEF-vdW exchange correlation functional. Our analysis suggests that all active sites of interest under operating conditions are oxygen saturated.

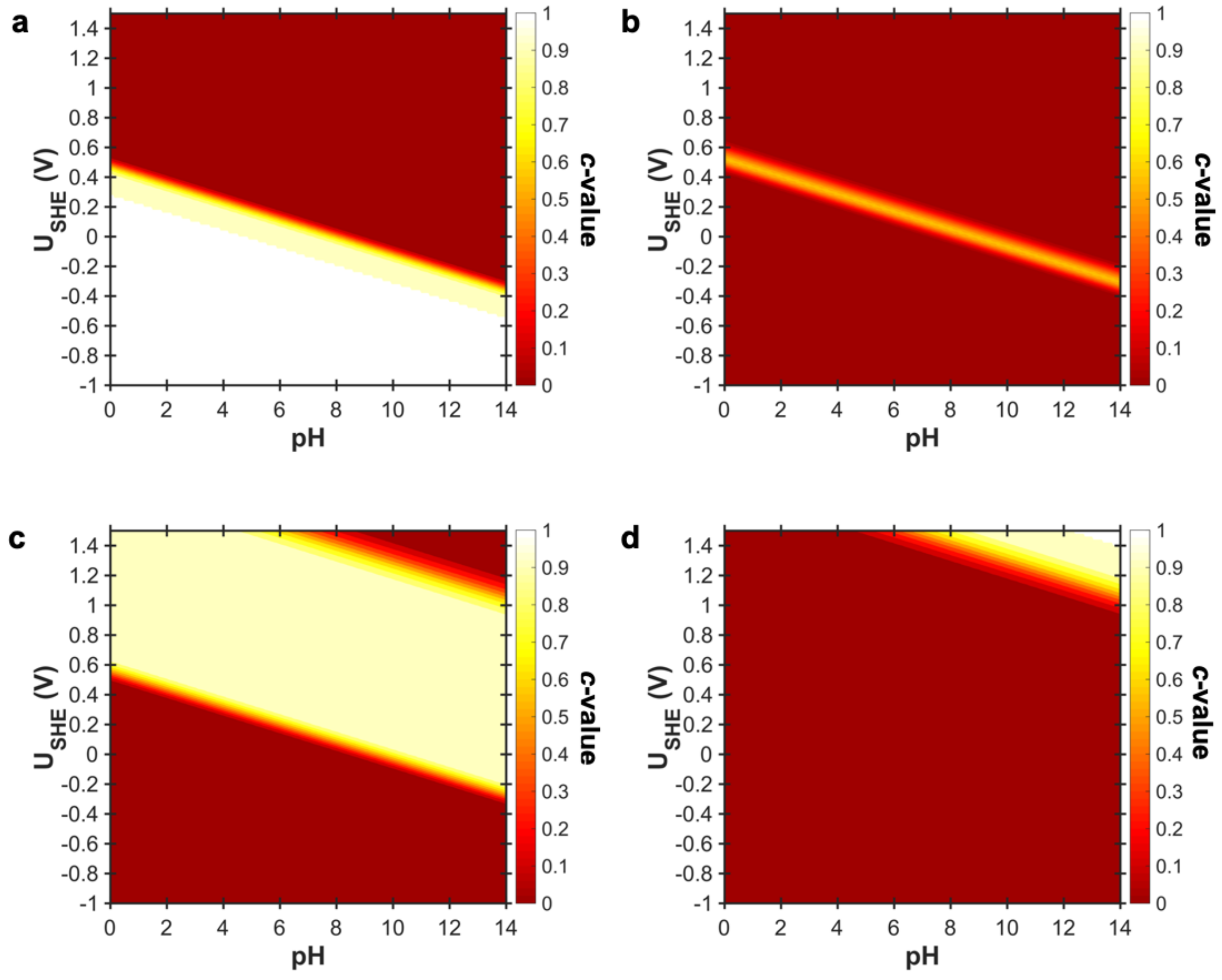

**Figure S21. Confidence values associated with predictions of the surface state of AC edge sites through a Pourbaix diagram.** Confidence values shown correspond to the surface states corresponding to **(a)** $H^*$, **(b)** $\frac{2}{3}OH^*\frac{1}{3}H^*$, **(c)** $\frac{2}{3}O^*\frac{1}{3}H^*$, and **(d)** $O^*$, which appear in the Pourbaix diagram as predicted using the optimal BEEF-vdW exchange correlation functional. The color bar indicates the associated confidence value (c-value) with the predicted surface state, which quantifies the degree of agreement between exchange correlation functionals at the chosen level of DFT complexity.

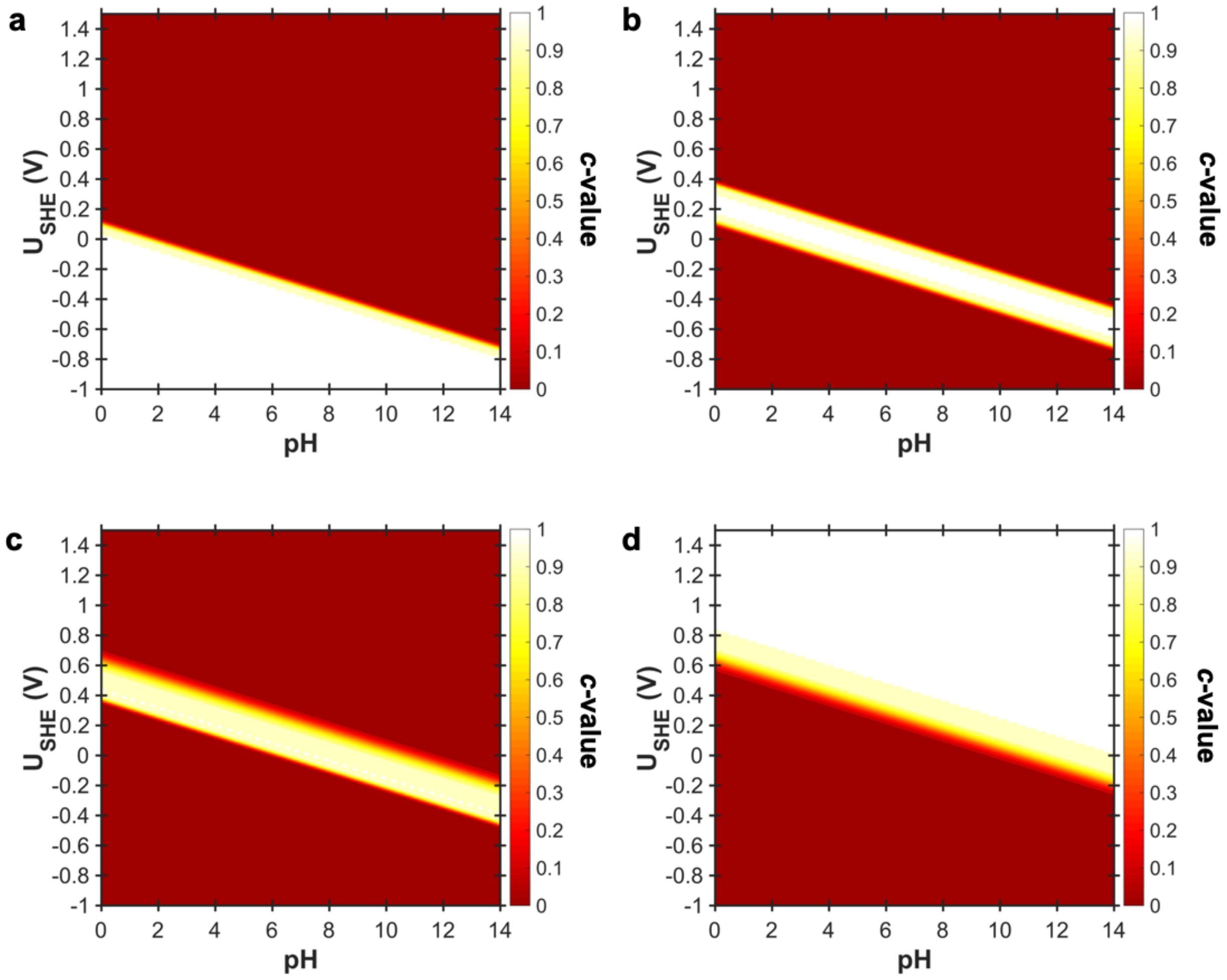

**Figure S22. Confidence values associated with predictions of the surface state of ZZ edge sites through a Pourbaix diagram.** Confidence values shown correspond to the surface states corresponding to **(a)** H*, **(b)** $\frac{1}{4}$O*$\frac{3}{4}$H*, **(c)** $\frac{1}{2}$O*$\frac{1}{2}$H*, and **(d)** O*, which appear in the Pourbaix diagram as predicted using the optimal BEEF-vdW exchange correlation functional. The color bar indicates the associated confidence value (c-value) with the predicted surface state, which quantifies the degree of agreement between exchange correlation functionals at the chosen level of DFT complexity.

| **Table S1. Data summary for Raman analysis for all NT-3DFG synthesis conditions.** Results are presented as mean ± SD (*n* = 3). | | | | | | | | |
|---|---|---|---|---|---|---|---|---|
| **Sample** | **$I_D$ $I_G^{-1}$** | **$I_{2D}$ $I_G^{-1}$** | **$Position_D$ ($cm^{-1}$)** | **$Position_G$ ($cm^{-1}$)** | **$Position_{2D}$ ($cm^{-1}$)** | **$FWHM_D$ ($cm^{-1}$)** | **$FWHM_G$ ($cm^{-1}$)** | **$FWHM_{2D}$ ($cm^{-1}$)** |
| **700 °C 30 min** | 2.34 ± 0.52 | 0.51 ± 0.09 | 1335.6 ± 3.8 | 1567.6 ± 5.4 | 2671.6 ± 7.4 | 48.8 ± 1.8 | 40.8 ± 2.2 | 83.4 ± 5.2 |
| **900 °C 30 min** | 3.59 ± 0.15 | 0.42 ± 0.03 | 1338.2 ± 1.2 | 1577.4 ± 2.4 | 2675.6 ± 3.8 | 58.0 ± 1.0 | 47.2 ± 1.6 | 102.4 ± 4.6 |
| **1100 °C 30 min** | 2.79 ± 0.11 | 0.52 ± 0.03 | 1338.2 ± 0.8 | 1579.0 ± 1.6 | 2674.6 ± 1.6 | 57.8 ± 1.0 | 49.0 ± 1.4 | 99.0 ± 4.0 |
| **1100 °C 60 min** | 2.36 ± 0.43 | 0.62 ± 0.13 | 1333.4 ± 2.6 | 1572.0 ± 4.4 | 2669.4 ± 5.0 | 57.6 ± 2.6 | 60.6 ± 3.0 | 99.6 ± 5.6 |
| **1100 °C 120 min** | 2.53 ± 0.16 | 0.61 ± 0.06 | 1338.2 ± 1.2 | 1578.6 ± 1.8 | 2676.6 ± 2.6 | 57.2 ± 1.4 | 57.6 ± 1.8 | 98.8 ± 5.0 |

| **Table S2. Data summary for Dual-laser Raman analysis of NT-3DFG synthesized at 700 °C for 30 min.** | | | | |
|---|---|---|---|---|
| **Sample** | **Region** | **$I_D$ $I_G^{-1}$** | **$FWHM_G$ ($cm^{-1}$)** | **$Disp_G$ ($cm^{-1}$ $nm^{-1}$)** |
| 1 | 1 | 2.46 | 34.2 | 0.12 |
| | 2 | 1.86 | 39.2 | 0.02 |
| | 3 | 2.11 | 37.8 | 0.02 |
| | 4 | 1.97 | 39.8 | 0.01 |
| | 5 | 2.73 | 39.6 | 0.03 |
| | 6 | 2.22 | 40.6 | 0.01 |
| | 7 | 1.95 | 40.2 | 0.02 |
| | 8 | 2.80 | 41.8 | 0.07 |
| | 9 | 2.43 | 39.6 | 0.02 |
| | 10 | 2.18 | 39.8 | 0.03 |
| 2 | 1 | 2.19 | 42.6 | 0.04 |
| | 2 | 2.55 | 42.4 | 0.03 |
| | 3 | 2.66 | 41.0 | 0.00 |
| | 4 | 2.89 | 43.4 | 0.00 |
| | 5 | 2.32 | 38.4 | 0.06 |
| | 6 | 2.00 | 41.4 | 0.03 |
| | 7 | 2.30 | 42.6 | 0.00 |
| | 8 | 2.21 | 42.0 | 0.00 |
| | 9 | 2.63 | 42.0 | 0.08 |
| | 10 | 2.70 | 43.6 | 0.05 |
| 3 | 1 | 2.16 | 39.6 | 0.03 |
| | 2 | 2.56 | 43.0 | 0.03 |
| | 3 | 1.96 | 39.6 | 0.01 |
| | 4 | 2.77 | 42.4 | 0.05 |
| | 5 | 2.05 | 41.2 | 0.06 |
| | 6 | 2.75 | 42.0 | 0.06 |
| | 7 | 2.19 | 40.4 | 0.02 |
| | 8 | 2.27 | 40.6 | 0.02 |
| | 9 | 2.24 | 38.8 | 0.00 |
| | 10 | 2.06 | 41.0 | 0.02 |

| Table S3. Data summary for Dual-laser Raman analysis of NT-3DFG synthesized at 900 °C for 30 min. | | | | |
|---|---|---|---|---|
| **Sample** | **Region** | **$I_D$ $I_G^{-1}$** | **$FWHM_G$ ($cm^{-1}$)** | **$Disp_G$ ($cm^{-1}$ $nm^{-1}$)** |
| 1 | 1 | 3.54 | 47.4 | 0.04 |
| | 2 | 3.63 | 46.8 | 0.08 |
| | 3 | 3.49 | 45.8 | 0.10 |
| | 4 | 3.49 | 47.8 | 0.05 |
| | 5 | 3.59 | 46.2 | 0.08 |
| | 6 | 3.56 | 47.4 | 0.07 |
| | 7 | 3.74 | 47.0 | 0.12 |
| | 8 | 3.57 | 46.2 | 0.05 |
| | 9 | 3.65 | 47.0 | 0.07 |
| | 10 | 3.51 | 46.4 | 0.04 |
| 2 | 1 | 3.54 | 46.0 | 0.04 |
| | 2 | 3.67 | 46.2 | 0.01 |
| | 3 | 3.72 | 48.0 | 0.06 |
| | 4 | 3.72 | 48.2 | 0.05 |
| | 5 | 3.59 | 47.8 | 0.06 |
| | 6 | 3.63 | 46.8 | 0.05 |
| | 7 | 3.51 | 46.8 | 0.06 |
| | 8 | 3.67 | 47.4 | 0.07 |
| | 9 | 3.55 | 47.0 | 0.05 |
| | 10 | 3.53 | 47.0 | 0.02 |
| 3 | 1 | 3.53 | 46.8 | 0.07 |
| | 2 | 3.44 | 48.8 | 0.00 |
| | 3 | 3.54 | 47.6 | 0.05 |
| | 4 | 3.71 | 47.0 | 0.11 |
| | 5 | 3.65 | 49.2 | 0.02 |
| | 6 | 3.53 | 48.8 | 0.18 |
| | 7 | 3.66 | 48.8 | 0.02 |
| | 8 | 3.54 | 47.0 | 0.09 |
| | 9 | 3.45 | 44.8 | 0.08 |
| | 10 | 3.73 | 48.4 | 0.14 |

| Table S4. Data summary for Dual-laser Raman analysis of NT-3DFG synthesized at 1100 °C for 30 min. | | | | |
|---|---|---|---|---|
| **Sample** | **Region** | **$I_D$ $I_G^{-1}$** | **$FWHM_G$ ($cm^{-1}$)** | **$Disp_G$ ($cm^{-1}$ $nm^{-1}$)** |
| 1 | 1 | 2.90 | 48.4 | 0.07 |
| | 2 | 2.86 | 48.2 | 0.05 |
| | 3 | 2.78 | 48.4 | 0.04 |
| | 4 | 2.82 | 48.8 | 0.05 |
| | 5 | 2.77 | 49.4 | 0.06 |
| | 6 | 2.91 | 48.6 | 0.03 |
| | 7 | 2.89 | 49.6 | 0.05 |
| | 8 | 2.90 | 47.6 | 0.04 |
| | 9 | 2.97 | 48.4 | 0.06 |
| | 10 | 2.86 | 49.4 | 0.04 |
| 2 | 1 | 2.81 | 48.8 | 0.07 |
| | 2 | 2.91 | 48.6 | 0.04 |
| | 3 | 2.91 | 50.2 | 0.08 |
| | 4 | 2.93 | 49.8 | 0.02 |
| | 5 | 2.83 | 50.0 | 0.09 |
| | 6 | 2.90 | 48.6 | 0.03 |
| | 7 | 2.73 | 46.8 | 0.07 |
| | 8 | 2.85 | 48.4 | 0.04 |
| | 9 | 2.87 | 48.4 | 0.06 |
| | 10 | 2.80 | 48.8 | 0.04 |
| 3 | 1 | 2.62 | 49.4 | 0.08 |
| | 2 | 2.66 | 50.0 | 0.11 |
| | 3 | 2.66 | 49.8 | 0.08 |
| | 4 | 2.64 | 49.6 | 0.05 |
| | 5 | 2.78 | 49.8 | 0.07 |
| | 6 | 2.67 | 49.4 | 0.05 |
| | 7 | 2.75 | 49.2 | 0.06 |
| | 8 | 2.61 | 50.8 | 0.03 |
| | 9 | 2.65 | 49.6 | 0.05 |
| | 10 | 2.60 | 48.6 | 0.06 |

| **Table S5. Data summary for Dual-laser Raman analysis of NT-3DFG synthesized at 1100 °C for 60 min.** | | | | |
|---|---|---|---|---|
| **Sample** | **Region** | **$I_D$ $I_G^{-1}$** | **$FWHM_G$ ($cm^{-1}$)** | **$Disp_G$ ($cm^{-1}$ $nm^{-1}$)** |
| 1 | 1 | 3.16 | 55.6 | 0.01 |
| | 2 | 2.75 | 57.2 | 0.03 |
| | 3 | 2.65 | 55.2 | 0.00 |
| | 4 | 2.99 | 55.6 | 0.01 |
| | 5 | 2.83 | 54.4 | 0.03 |
| | 6 | 2.75 | 54.4 | 0.03 |
| | 7 | 2.77 | 55.6 | 0.01 |
| | 8 | 2.73 | 55.0 | 0.03 |
| | 9 | 2.83 | 58.6 | 0.01 |
| | 10 | 2.72 | 54.8 | 0.02 |
| 2 | 1 | 2.17 | 61.6 | 0.00 |
| | 2 | 2.10 | 65.0 | 0.02 |
| | 3 | 2.07 | 62.8 | 0.01 |
| | 4 | 1.77 | 62.4 | 0.00 |
| | 5 | 1.88 | 63.4 | 0.06 |
| | 6 | 2.20 | 60.0 | 0.00 |
| | 7 | 2.03 | 59.6 | 0.03 |
| | 8 | 2.08 | 63.6 | 0.02 |
| | 9 | 2.14 | 61.0 | 0.01 |
| | 10 | 2.10 | 60.8 | 0.01 |
| 3 | 1 | 1.92 | 65.2 | 0.04 |
| | 2 | 2.15 | 64.6 | 0.05 |
| | 3 | 3.15 | 68.8 | 0.02 |
| | 4 | 1.97 | 62.4 | 0.03 |
| | 5 | 2.56 | 65.8 | 0.06 |
| | 6 | 1.86 | 62.8 | 0.00 |
| | 7 | 2.17 | 60.8 | 0.02 |
| | 8 | 2.05 | 64.8 | 0.07 |
| | 9 | 2.15 | 65.0 | 0.08 |
| | 10 | 2.08 | 63.6 | 0.00 |

| **Table S6. Data summary for Dual-laser Raman analysis of NT-3DFG synthesized at 1100 °C for 120 min.** | | | | |
|---|---|---|---|---|
| **Sample** | **Region** | **$I_D$ $I_G^{-1}$** | **$FWHM_G$ ($cm^{-1}$)** | **$Disp_G$ ($cm^{-1}$ $nm^{-1}$)** |
| 1 | 1 | 2.23 | 62.2 | 0.08 |
| | 2 | 2.37 | 64.8 | 0.09 |
| | 3 | 2.38 | 66.0 | 0.10 |
| | 4 | 2.04 | 63.6 | 0.07 |
| | 5 | 2.15 | 61.2 | 0.08 |
| | 6 | 2.20 | 63.4 | 0.05 |
| | 7 | 2.21 | 62.8 | 0.05 |
| | 8 | 2.38 | 64.6 | 0.08 |
| | 9 | 2.44 | 65.2 | 0.02 |
| | 10 | 2.18 | 64.8 | 0.10 |
| 2 | 1 | 2.57 | 55.6 | 0.04 |
| | 2 | 2.47 | 54.2 | 0.03 |
| | 3 | 2.54 | 55.2 | 0.02 |
| | 4 | 2.58 | 57.4 | 0.01 |
| | 5 | 2.35 | 55.2 | 0.00 |
| | 6 | 2.38 | 54.2 | 0.04 |
| | 7 | 2.50 | 55.4 | 0.01 |
| | 8 | 2.58 | 56.0 | 0.01 |
| | 9 | 2.65 | 55.4 | 0.04 |
| | 10 | 2.47 | 55.2 | 0.00 |
| 3 | 1 | 2.78 | 54.2 | 0.06 |
| | 2 | 2.81 | 52.6 | 0.05 |
| | 3 | 2.82 | 52.6 | 0.03 |
| | 4 | 2.82 | 53.4 | 0.03 |
| | 5 | 2.80 | 53.6 | 0.04 |
| | 6 | 2.85 | 53.4 | 0.03 |
| | 7 | 2.86 | 52.6 | 0.03 |
| | 8 | 2.81 | 53.2 | 0.03 |
| | 9 | 2.93 | 54.0 | 0.03 |
| | 10 | 2.81 | 51.8 | 0.02 |

**Table S7. Data Summary for double layer capacitance characterization.** Results presented as mean ± SD ($n$ = 3). Specific capacitance was estimated as 20 µF $cm^2_{ECSA}$.

| | **Capacitance ($\mu F/cm^2_{geo.}$)** | | **Electrochemical Surface Area ($cm^2_{ECSA}/cm^2_{geo.}$)** | |
|---|---|---|---|---|
| | **µ** | **σ** | **µ** | **σ** |
| **700 °C 30 min** | **102.2** | **5.1** | **6.8** | **0.3** |
| **900 °C 30 min** | **954.8** | **101.3** | **63.7** | **6.8** |
| **1100 °C 30 min** | **1084.2** | **235.6** | **72.3** | **15.7** |
| **1100 °C 60 min** | **2934.4** | **518.1** | **195.6** | **34.5** |
| **1100 °C 120 min** | **4361.9** | **1103.1** | **290.8** | **73.5** |

**Table S8. Data Summary for electrical characterization.** Results presented as mean ± SD ($n$ = 3). Resistance contribution to circuit by electronic conduction through catalyst layer, $R_{electrode}$ was calculated as $R_{\text{electrode}} = \frac{\rho_{\text{electrode}} L_{\text{thickness}}}{A_{\text{crosssection}}}$ where $L_{thickness}$ was the thickness of the catalyst layer as measured via SEM imaging.

| **Condition** | **Sheet resistance (Ω/□)** | | **Mesh Thickness (µm)** | | **Resistivity (Ω*cm)** | | **Calculated $iR_{electrode}$ loss (V)** |
|---|---|---|---|---|---|---|---|
| | **µ** | **σ** | **µ** | **σ** | **µ** | **σ** | **$i$ = 0.5 mA, A = 0.196 $cm^2$** |
| **700 °C 30 min** | **311.0** | **41.2** | **11.21** | **1.08** | **0.349** | **0.080** | **8.90E-07** |
| **1100 °C 30 min** | **162.7** | **35.4** | **10.49** | **1.60** | **0.171** | **0.063** | **4.35E-07** |
| **1100 °C 120 min** | **45.3** | **15.1** | **15.17** | **0.94** | **0.069** | **0.027** | **1.75E-07** |

| **Table S9. Data Summary for electrical impedance spectroscopy.** EIS was performed at -0.2 V versus Ag/AgCl in $O_2$-saturated 0.1 M KOH. The Nyquist plots for each EIS spectra were fit with an equivalent Randles circuit (**Figure 3.b inset**) using PSTrace software. | | | | |
|---|---|---|---|---|
| Sample | **$R_s$ (Ω)** | **$R_{ct}$ (Ω)** | **$C_{dl}$ (μF)** | **Goodness of Fit ($X^2$)** |
| 700 °C 30 min, 1 | 62 | 4588 | 38 | 0.092 |
| 700 °C 30 min, 2 | 62 | 3388 | 45 | 0.078 |
| 700 °C 30 min, 3 | 56 | 3633 | 54 | 0.086 |
| **Average** | **60** | **3870** | **46** | |
| **Stddev** | **3** | **634** | **8** | |
| 1100 °C 30 min, 1 | 59 | 2067 | 217 | 0.075 |
| 1100 °C 30 min, 2 | 73 | 1690 | 200 | 0.064 |
| 1100 °C 30 min, 3 | 57 | 2197 | 233 | 0.085 |
| **Average** | **63** | **1985** | **216** | |
| **Stddev** | **9** | **263** | **17** | |
| 1100 °C 120 min, 1 | 71 | 913 | 404 | 0.091 |
| 1100 °C 120 min, 2 | 59 | 1263 | 466 | 0.080 |
| 1100 °C 120 min, 3 | 67 | 1727 | 479 | 0.110 |
| **Average** | **65** | **1301** | **450** | |
| **Stddev** | **6** | **408** | **40** | |
| **GC** | 58 | 45840 | 8 | 0.123 |

| Table S10. Data Summary for $H_2O_2$ Selectivity of NT-3DFG. | | | | | |
|---|---|---|---|---|---|
| | Charge transferred (C) | e- (μmol) | $H_2O_2$ (μmol) | n | Selectivity (%) |
| 700 °C 30 min, 1 | -0.199 | 2.059 | 0.940 | 2.191 | 90.5 |
| 700 °C 30 min, 2 | -0.178 | 1.848 | 0.780 | 2.370 | 81.5 |
| 700 °C 30 min, 3 | -0.196 | 2.034 | 0.880 | 2.312 | 84.4 |
| **Average** | **-0.191** | **1.981** | **0.867** | **2.291** | **85.5** |
| **Stddev** | **0.011** | **0.115** | **0.081** | **0.091** | **4.6** |
| 1100 °C 30 min, 1 | -0.199 | 2.066 | 0.930 | 2.222 | 88.9 |
| 1100 °C 30 min, 2 | -0.254 | 2.631 | 1.190 | 2.211 | 89.5 |
| 1100 °C 30 min, 3 | -0.294 | 3.044 | 1.440 | 2.114 | 94.3 |
| **Average** | **-0.249** | **2.580** | **1.187** | **2.182** | **90.9** |
| **Stddev** | **0.047** | **0.491** | **0.255** | **0.060** | **3.0** |
| 1100 °C 120 min, 1 | -0.303 | 3.139 | 1.430 | 2.195 | 90.3 |
| 1100 °C 120 min, 2 | -0.397 | 4.117 | 1.950 | 2.111 | 94.4 |
| 1100 °C 120 min, 3 | -0.305 | 3.158 | 1.510 | 2.091 | 95.4 |
| **Average** | **-0.335** | **3.471** | **1.630** | **2.132** | **93.4** |
| **Stddev** | **0.054** | **0.559** | **0.280** | **0.055** | **2.7** |

| Table S11. Data summary for Raman analysis for NT-3DFG before & after ORR. | | | | | | | | |
|---|---|---|---|---|---|---|---|---|
| Sample | $I_D\ I_G^{-1}$ | $I_{2D}\ I_G^{-1}$ | $Position_D$ ($cm^{-1}$) | $Position_G$ ($cm^{-1}$) | $Position_{2D}$ ($cm^{-1}$) | $FWHM_D$ ($cm^{-1}$) | $FWHM_G$ ($cm^{-1}$) | $FWHM_{2D}$ ($cm^{-1}$) |
| **Pristine** | 2.83 ± 0.04 | 0.53 ± 0.04 | 1335.8 ± 0.6 | 1578.4 ± 0.8 | 2674.4 ± 1.2 | 55.0 ± 0.8 | 53.2 ± 0.8 | 98.6 ± 2.2 |
| **Post-ORR** | 2.84 ± 0.13 | 0.48 ± 0.04 | 1338.6 ± 0.8 | 1574.2 ± 1.8 | 2679.6 ± 2.2 | 60.6 ± 0.6 | 60.0 ± 1.8 | 102.4 ± 1.4 |

| **Table S12. Data summary for XPS analysis of NT-3DFG before & after ORR.** ORR testing included cyclic voltammetry and $H_2O_2$ selectivity. Results are presented as mean ± SD (*n* = 3). | | | | | |
|---|---|---|---|---|---|
| **Condition** | **Sample** | **Peak** | **Peak BE (eV)** | **Peak FWHM (eV)** | **Composition (at %)** |
| Pristine | *n* = 1 | C1s | 284.9 | 3.1 | 99.6 |
| | | O1s | 532.9 | - | trace (< 1) |
| | *n* = 2 | C1s | 285.1 | 3.4 | 99.1 |
| | | O1s | 533.2 | - | trace (< 1) |
| | *n* = 3 | C1s | 284.9 | 3.2 | 99.8 |
| | | O1s | 531.7 | - | trace (< 1) |
| | average | C1s | **285.0 ± 0.1** | **3.2 ± 0.1** | **99.5 ± 0.4** |
| | | O1s | **532.6 ± 0.8** | **-** | **-** |
| | | | | | |
| Post ORR | *n* = 1 | C1s | 284.9 | 3.5 | 86.1 |
| | | O1s | 533.1 | 3.2 | 13.2 |
| | | Si2s | 153.5 | - | trace (< 1) |
| | *n* = 2 | C1s | 284.7 | 2.6 | 87.6 |
| | | O1s | 532.6 | 3.4 | 10.7 |
| | | Si2s | 153.1 | 3.7 | 1.7 |
| | *n* = 3 | C1s | 284.7 | 2.7 | 88.6 |
| | | O1s | 532.6 | 3.6 | 10.5 |
| | | Si2s | 152.6 | - | trace (< 1) |
| | | N1s | 399.6 | - | trace (< 1) |
| | average | C1s | **284.8 ± 0.1** | **2.9 ± 0.5** | **87.4 ± 1.2** |
| | | O1s | **532.8 ± 0.3** | **3.4 ± 0.2** | **11.5 ± 1.5** |
| | | Si2s | **153.1 ± 0.5** | **-** | **-** |

**Table S13. Data summary for C1s & O1s chemical states of NT-3DFG after ORR.** ORR testing included cyclic voltammetry and $H_2O_2$ selectivity. Results presented as mean ± SD (*n* = 3).

| Peak | Sample | State | Peak BE (eV) | FWHM (eV) | Ratio |
|---|---|---|---|---|---|
| C1s | *n* = 1 | C-C | 284.3 | 1.3 | 90.9 |
| | | C-O | 286.0 | 1.4 | 4.5 |
| | | C=O | 287.0 | 1.4 | 1.8 |
| | | O=C-O | 288.5 | 1.4 | 2.7 |
| | | Plasmon | 290.6 | 2.0 | - |
| | *n* = 2 | C-C | 284.1 | 1.2 | 83.3 |
| | | C-O | 285.9 | 1.4 | 9.2 |
| | | C=O | 286.9 | 1.4 | 5.0 |
| | | O=C-O | 288.1 | 1.4 | 2.5 |
| | | Plasmon | 290.5 | 3.7 | - |
| | *n* = 3 | C-C | 283.9 | 1.1 | 87.0 |
| | | C-O | 285.2 | 1.4 | 7.8 |
| | | C=O | 286.4 | 1.4 | 2.6 |
| | | O=C-O | 287.7 | 1.4 | 2.6 |
| | | Plasmon | 290.2 | 3.0 | - |
| | average | C-C | **284.1 ± 0.2** | **1.2 ± 0.1** | **87.1 ± 3.8** |
| | | C-O | **285.7 ± 0.4** | **1.4 ± 0.1** | **7.2 ± 2.4** |
| | | C=O | **286.8 ± 0.3** | **1.4 ± 0.1** | **3.1 ± 1.7** |
| | | O=C-O | **288.1 ± 0.4** | **1.4 ± 0.1** | **2.6 ± 0.1** |
| | | Plasmon | **290.4 ± 0.2** | **2.9 ± 0.9** | **-** |
| O1s | *n* = 1 | C=O | 531.5 | 1.8 | 49.8 |
| | | C-O | 532.9 | 1.8 | 50.2 |
| | *n* = 2 | C=O | 531.5 | 1.8 | 46.4 |
| | | C-O | 533.0 | 1.8 | 53.6 |
| | *n* = 3 | C=O | 531.9 | 1.8 | 38.0 |
| | | C-O | 532.9 | 1.8 | 62.0 |
| | average | C=O | **531.6 ± 0.2** | **1.8 ± 0.1** | **44.7 ± 6.1** |
| | | C-O | **532.9 ± 0.1** | **1.8 ± 0.1** | **55.3 ± 6.1** |

**Table S14. Data summary for explored ZZ and AC active sites including local coordination environment.**

| Active Site | | $\Delta G_{OOH}$ | $N_O$ | $N_{OH}$ | $N_H$ | $N_C$ | $N_{C\text{-}O}$ | $N_{C\text{-}OH}$ | $N_{C\text{-}H}$ |
|---|---|---|---|---|---|---|---|---|---|
| ZZ | H* | 3.914 | 0 | 0 | 1 | 2 | 0 | 0 | 0 |
| | H* (basal) | 5.308 | 0 | 0 | 0 | 3 | 0 | 0 | 0 |
| | H* | 3.905 | 0 | 0 | 1 | 2 | 0 | 0 | 0 |
| | 3/4O*1/4H* | 4.369 | 0 | 0 | 0 | 1 | 2 | 0 | 0 |
| | O* | 4.805 | 0 | 0 | 0 | 1 | 2 | 0 | 0 |
| | 1/4OH*3/4H* | 3.890 | 0 | 1 | 0 | 2 | 0 | 0 | 0 |
| | OH* | 3.868 | 0 | 1 | 0 | 2 | 0 | 0 | 0 |
| | 1/4O*3/4H* | 5.411 | 0 | 0 | 1 | 3 | 0 | 0 | 0 |
| | 3/4O*1/4H* | 4.663 | 0 | 0 | 0 | 1 | 2 | 0 | 0 |
| | 1/4O*3/4H* | 5.342 | 0 | 0 | 0 | 1 | 1 | 0 | 1 |
| | 1/2O*1/2H* | 5.252 | 0 | 0 | 0 | 1 | 1 | 0 | 1 |
| | 3/4O*1/4H* | 4.369 | 0 | 0 | 0 | 1 | 2 | 0 | 0 |
| | 1/2OH*1/2H* | 3.902 | 0 | 1 | 0 | 2 | 0 | 0 | 0 |
| | 1/2OH*1/2H* | 4.935 | 0 | 0 | 0 | 1 | 0 | 1 | 1 |
| | 3/4OH*1/4H* | 3.893 | 0 | 1 | 0 | 2 | 0 | 0 | 0 |
| | 3/4OH*1/4H* | 3.811 | 0 | 1 | 0 | 2 | 0 | 0 | 0 |
| | 3/4OH*1/4H* | 3.682 | 0 | 0 | 1 | 2 | 0 | 0 | 0 |
| | OH* | 4.812 | 0 | 0 | 0 | 1 | 0 | 2 | 0 |
| AC | 1/6O*5/6H* | 4.029 | 0 | 0 | 1 | 1 | 1 | 0 | 0 |
| | 1/3O*2/3H* | 3.815 | 0 | 0 | 1 | 1 | 1 | 0 | 0 |
| | 1/2O*1/2H* | 3.626 | 0 | 0 | 1 | 1 | 1 | 0 | 0 |
| | 2/3O*1/3H* | 3.808 | 0 | 0 | 1 | 1 | 1 | 0 | 0 |
| | OH* | 3.886 | 0 | 1 | 0 | 1 | 0 | 1 | 0 |
| | 1/3O*2/3H* | 4.633 | 0 | 0 | 0 | 2 | 1 | 0 | 0 |
| | 1/3O*2/3H* | 3.975 | 0 | 0 | 1 | 1 | 1 | 0 | 0 |
| | 1/2O*1/2H* | 3.644 | 0 | 0 | 1 | 1 | 1 | 0 | 0 |
| | 1/6O*5/6H* | 4.906 | 0 | 0 | 0 | 2 | 1 | 0 | 0 |
| | 1/6O*5/6H* | 4.048 | 0 | 0 | 1 | 1 | 1 | 0 | 0 |
| | 1/2OH*1/2H* | 4.491 | 0 | 1 | 0 | 1 | 0 | 0 | 1 |